\documentclass{article}
\usepackage{amsmath}
\usepackage{graphicx}
\usepackage{amsfonts}
\usepackage{amssymb}
\renewcommand{\theequation}{\arabic{section}.\arabic{equation}} 
\begin{document}

\title{Inequalities for Wilson loops, cusp singularities, area law and shape of a drum}
\author{P.V. Pobylitsa}
\date{}
\maketitle

\begin{center}
\emph{Institute for Theoretical Physics II, Ruhr University Bochum,
\\ D-44780 Bochum, Germany }\\ and \\ \emph{Petersburg
Nuclear Physics Institute, Gatchina, \\ St. Petersburg, 188300, Russia}
\end{center}

\begin{abstract}
Inequalities are derived for Wilson loops generalizing the well-known Bachas
inequality for rectangular contours. The inequalities are compatible with the
area law for large contours. The Polyakov cusp anomalous dimension of Wilson
lines (playing an important role in QCD applications to hard processes) has a
convex angular dependence. This convexity is crucial for the consistency of
the inequalities with renormalization. Some parallel properties can be found
in the string theory. The Kac-Ray cusp term from the ``shape of a drum"
problem has the same angular convexity property and plays the role of the cusp
anomalous dimension in the effective string model for Wilson loops studied by
L\"{u}scher, Symanzik and Weisz (LSW). Using heuristic arguments based on the
LSW model, one can find an interesting connection between the inequalities for
Wilson loops and inequalities for determinants of two-dimensional Laplacians
with Dirichlet boundary conditions on the closed contours associated with
Wilson loops.

\end{abstract}

\section{Introduction}

\setcounter{equation}{0} 

\subsection{Inequalities for rectangular Wilson loops and properties of static
heavy-quark potential}

Among various inequalities for Wilson loops \cite{Wilson-74},
\begin{equation}
W(C)=\frac{1}{N}\left\langle \mathrm{Tr\,P}\left(  \exp i\oint_{C}A\,dx\right)
\right\rangle \,, \label{W-C-general}\end{equation}
an important role is played by Bachas inequality \cite{Bachas-86} (see also
\cite{Seiler-1978,BS-83}) for \emph{rectangular} $T\times R$ Wilson loops
$W(T,R)$
\begin{equation}
\left[  W\left(  T,\frac{R_{1}+R_{2}}{2}\right)  \right]  ^{2}\leq W\left(
T,R_{1}\right)  W\left(  T,R_{2}\right)  \,. \label{Bachas-ineq}
\end{equation}
This inequality was derived in the lattice gauge theory using the reflection
positivity property \cite{OS-75,OS-78,BFS-78} (which corresponds to the
positive metric of the Hilbert space for physical states in the operator
formulation of the theory). The large $T$ behavior of $\ln W\left(
T,R\right)  $ is described by the static heavy-quark potential
\begin{equation}
V(R)=-\lim_{T\rightarrow\infty}\frac{1}{T}\ln W\left(  T,R\right)  \,.
\label{V-via-W}
\end{equation}
In Ref. \cite{Bachas-86} inequality (\ref{Bachas-ineq}) was used in order to
derive the convexity of the potential
\begin{equation}
V\left(  \frac{R_{1}+R_{2}}{2}\right)  \geq\frac{1}{2}
\left[  V(R_{1})+V(R_{2})\,\right]  .\label{V-convexity}
\end{equation}
In the continuum limit, this convexity property is equivalent to the
inequality
\begin{equation}
\frac{d^{2}V(R)}{dR^{2}}\leq0\,.\label{V-sec-der-negative}
\end{equation}
This inequality has several important consequences. For example, it shows that
the potential $V(R)$ cannot grow faster than linearly, the large distance
behavior $V(R)\sim R^{\alpha}$ with $\alpha>1$ is forbidden. Another
consequence is that the $1/R$ correction to the linear potential $KR$ can be
only negative
\begin{gather}
V(R)\overset{R\rightarrow\infty}{=}KR+2a_{0}+a_{-1}\frac{1}{R}+\ldots
\,,\label{V-large-R}\\
a_{-1}\leq0\,.
\end{gather}

\subsection{Cusp anomalous dimension and its convexity}

\begin{figure}
[ptb]
\begin{center}
\includegraphics[
height=1.3292in,
width=1.1857in
]%
{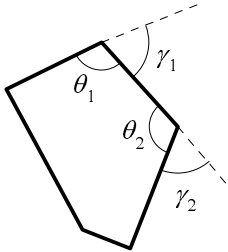}%
\caption{Cusp angles: inside-facing angles $\theta_{k}$ and deviation angles
$\gamma_{k}=\pi-\theta_{k}$.}%
\label{fig-1}%
\end{center}
\end{figure}

In Ref. \cite{Polyakov-79} A.M. Polyakov has shown that the renormalization of
Wilson loops corresponding to non-smooth contours is accompanied by additional
renormalization factors depending on the cusp angles\footnote{Following the
tradition we write cusp anomalous dimension (\ref{Gamma-1-cusp}) \emph{in the
gauge theory} as a function of the deviation angle $\gamma=\pi-\theta$ where
$\theta$ is the inside-facing angle (see Fig.~\ref{fig-1}). However, \emph{in
the effective string model} discussed in Sec.~\ref{String-section} we use
angle $\theta$.} $\gamma$ (Fig \ref{fig-1}). These renormalization factors can
be interpreted in terms of the cusp anomalous dimension 
$\Gamma_{\mathrm{cusp}}(\gamma)$. To the leading order\footnote{The NLO correction was computed in
Ref. \cite{KR-87}.} in the coupling $g$, this anomalous dimension for the
SU($N_{c}$) gauge group is given by
\begin{equation}
\Gamma_{\mathrm{cusp}}^{(1)}(\gamma)=-\frac{g^{2}}{4\pi^{2}}\frac{N_{c}^{2}
-1}{2N_{c}}\left(  1-\gamma\cot\gamma\right)  \,.\label{Gamma-1-cusp}
\end{equation}
A straightforward check shows that $\Gamma_{\mathrm{cusp}}^{(1)}(\gamma)$ has
the convexity property (see Sec.~\ref{convexity-AD-subsection} for details)
\begin{equation}
\Gamma_{\mathrm{cusp}}^{(1)}\left(  \frac{\gamma_{1}+\gamma_{2}}{2}\right)
\geq\frac{1}{2}\left[  \Gamma_{\mathrm{cusp}}^{(1)}(\gamma_{1})+\Gamma
_{\mathrm{cusp}}^{(1)}(\gamma_{2})\,\right]  .\label{Gamma-1-convex}%
\end{equation}

\subsection{Polygonal Wilson loops as a common basis for perturbative and
nonperturbative bounds}

At first sight, the formal coincidence of the convexity properties
(\ref{V-convexity}), (\ref{Gamma-1-convex}) for the static potential $V(R)$
and for the cusp anomalous dimension $\Gamma_{\mathrm{cusp}}^{(1)}(\gamma)$ is
a pure mathematical curiosity which has nothing to do with physics. The
heavy-quark potential $V(R)$ is a nonperturbative quantity describing both
large and short distances. The anomalous dimension $\Gamma_{\mathrm{cusp}%
}^{(1)}(\gamma)$ arises in the context of the perturbative renormalization and
plays an important role in the physics of small distances and hard processes
\cite{KR-87}. The convexity of the potential $V(R)$ was rigorously established
in the lattice gauge theory (with discrete $R$), whereas the cusp anomalous
dimension can be discussed only in the continuum limit.

Nevertheless it is possible to find an interesting connection between the
convexity of the potential $V(R)$ and the convexity of the anomalous dimension
$\Gamma_{\mathrm{cusp}}^{(1)}(\gamma)$. In this paper it will be shown that
both convexity properties naturally appear in a common framework provided by
inequalities for polygonal Wilson loops. The idea to consider polygonal Wilson
loops is quite natural. On the one hand, we already know that the convexity of
the static quark potential (\ref{V-convexity}) follows from the Bachas
inequality for rectangular Wilson loops \ref{Bachas-ineq}. On the other hand,
the renormalization of polygonal Wilson loops provides an access to the cusp
anomalous dimension.

Certainly, the interpretation of the convexity of the cusp anomalous dimension
is not the only motivation for studying inequalities for polygonal Wilson
loops. For example, one could be interested in the constraints imposed by the
inequalities on the area law behavior in confining theories.

\subsection{Example: trapezium-parallelogram inequality}

In this article we will derive inequalities for Wilson loops with rather
general contours. To begin, let us take a very simple, but still instructive example.

\begin{figure}
[ptb]
\begin{center}
\includegraphics[
height=3.3044in,
width=3.2768in
]
{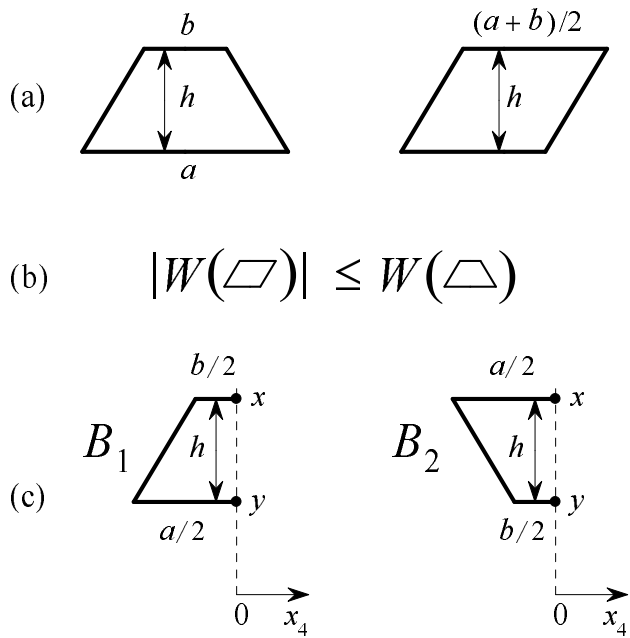}
\caption{(a) Trapezium and parallelogram entering in inequality
(\ref{parall-trap-ineq}). (b) Graphic representation of inequalty
(\ref{parall-trap-ineq}) for the Wilson loops. (c) Open contours $B_{1}$ and
$B_{2}$ used for the derivation of inequality (\ref{parall-trap-ineq}). The
end points obey the condition $x_{4}=y_{4}=0$.}%
\label{fig-2}%
\end{center}
\end{figure}

Let us consider a symmetric trapezium with the height $h$ and with two
parallel sides $a$ and $b$ as shown in Fig.~\ref{fig-2} (a). The corresponding
Wilson loop will be denoted $W_{\mathrm{trapezium}}(a,b,h)$. We use notation
$W_{\mathrm{parallelogram}}(c,h)$ for the Wilson loop associated with the
parallelogram with side $c$ and height $h$. In Sec.
\ref{Derivation-t-p-subsection} we will derive the inequality%
\begin{equation}
\left|  W_{\mathrm{parallelogram}}\left(  \frac{a+b}{2},h\right)  \right|
\leq W_{\mathrm{trapezium}}(a,b,h) \label{parall-trap-ineq}%
\end{equation}
which is shown graphically in Fig.~\ref{fig-2} (b).%

\subsection{Inequalities and renormalization}

Trapezium-parallelogram inequality (\ref{parall-trap-ineq}) is just a simple
example of the wide class of inequalities which are discussed in this article.
The derivation of inequalities for polygonal Wilson loops follows the same
ideas as the derivation of the Bachas inequality for rectangular contours
(\ref{Bachas-ineq}). However, in the case of polygonal contours one meets
several problems:

1) Rectangular loops can be studied in the context of lattice gauge theory. In
the case of general polygonal contours we have to work with the continuum
version of the theory.

2) The continuum limit is accompanied by renormalization. Although the
asymptotic freedom of non-Abelian gauge theories provides a rather reliable
\emph{perturbative control} of the renormalization of \emph{nonperturbative
quantities}, one often has to use heuristic arguments instead of
``mathematically rigorous'' statements of the lattices gauge theory.

3) Polygonal contours have cusps which give an additional contribution to the
renormalization of Wilson loops. In the case of rectangular contours the cusp
renormalization factor (determined by the four 90${{}^\circ}$
vertices of the rectangle) is the same for all rectangles. On the contrary,
the renormalization of cusp singularities in polygonal Wilson loops is not
universal. Therefore a special care must be taken about the compatibility of
the naive inequalities with renormalization.

Our analysis of the renormalization of inequalities for Wilson loops leads to
the following conclusions:

1) One class of inequalities survives under renormalization. In these
inequalities, all renormalization constants cancel, so that the correct
renormalized form of the inequalities coincides with naive nonrenormalized version.

2) In the second class of inequalities, the balance of the renormalization
constants is violated. Therefore in the renormalized version of the inequality
some contributions of the original nonrenormalized inequality die out. As a
result, one arrives at trivial inequalities.

Although the trivial renormalized inequalities of the second class are not
interesting, the fact that these trivial inequalities are \emph{correct}
deserves some attention. This correctness comes as a result of a peculiar
property of cusp renormalization factors, which follows from the convexity of
the angular dependence of cusp anomalous dimension. Therefore we see that
there is a rather interesting connection between the general positivity
properties of the theory (underlying the inequality) and properties of the
cusp anomalous dimension.

Another interesting issue is the connection between the inequalities discussed
in this paper and the area law for Wilson loops expected in gauge theories
confining heavy quarks. Here we have the following situation: the area law
does not follow from the inequalities but all inequalities are compatible with
the area law.

\subsection{From gauge theory to determinants of two-dimensional Laplacians
via effective string model}

In the effective string model suggested by L\"{u}scher, Symanzik and Weisz
(LSW) \cite{Luescher-80}, \cite{Luescher-81} for the description of large
Wilson loops, contours with cusps also have additional ultraviolet
divergences. Similar to gauge theories, these divergences factorize into a
product of single cusped contributions. The dependence of the divergence on
the inside-facing cusp angle $\theta$ (see Fig.~\ref{fig-1}) is described by
the anomalous dimension%
\begin{equation}
\Gamma_{\mathrm{cusp}}^{\mathrm{str}}(\theta)=\frac{\theta^{2}-\pi^{2}}%
{24\pi\theta}\,. \label{Gamma-cusp-string}%
\end{equation}

The analogy with gauge theories can be continued. Function $\Gamma
_{\mathrm{cusp}}^{\mathrm{str}}(\theta)$ has the same convexity property%
\begin{equation}
\Gamma_{\mathrm{cusp}}^{\mathrm{str}}\left(  \frac{\theta_{1}+\theta_{2}}%
{2}\right)  \geq\frac{1}{2}\left[  \Gamma_{\mathrm{cusp}}^{\mathrm{str}%
}(\theta_{1})+\Gamma_{\mathrm{cusp}}^{\mathrm{str}}(\theta_{2})\,\right]
\quad(0<\theta<2\pi) \label{Gamma-1-convex-string}%
\end{equation}
as Polyakov's anomalous dimension (\ref{Gamma-1-cusp}), (\ref{Gamma-1-convex}).

Expression (\ref{Gamma-cusp-string}) for the cusp anomalous dimension in the
string model has an old history. M. Kac has studied the spectrum of the
two-dimensional Laplace operator $\Delta_{C}$ with Dirichlet conditions on the
boundary $C$ using the trace of the operator $e^{t\Delta_{C}}$. In paper ``Can
one hear the shape of a drum?'' \cite{Kac-1966} he constructed the small-$t$
expansion of $\mathrm{Tr}\,e^{t\Delta_{C}}$ for polygonal contours $C$ with
cusp angles $\theta_{i}$:%

\begin{equation}
\mathrm{Tr}\,e^{t\Delta_{C}}=t^{-1}\frac{S(C)}{4\pi}-t^{-1/2}\frac{L(C)}%
{8\sqrt{\pi}}-t^{0}\sum\limits_{i}\Gamma_{\mathrm{cusp}}^{\mathrm{str}}%
(\theta_{i})+\ldots\label{sp-exp-expansion}%
\end{equation}
Here $S(C)$ is the surface area bounded by the contour $C$ and $L(C)$ is the
length of $C$. The cusp function $\Gamma_{\mathrm{cusp}}^{\mathrm{str}}%
(\theta)$ obtained by M. Kac was represented by a complicated integral in Ref.
\cite{Kac-1966}. The simple expression (\ref{Gamma-cusp-string}) for
$\Gamma_{\mathrm{cusp}}^{\mathrm{str}}(\theta)$ was obtained by D.B. Ray (the
derivation is described in Ref. \cite{McKean-Singer-67}).

Inserting the small-$t$ expansion (\ref{sp-exp-expansion}) into the proper
time representation for determinants%
\begin{equation}
\ln\mathrm{Det}\,\frac{\Delta_{1}}{\Delta_{2}}=-\int_{0}^{\infty}\frac{dt}%
{t}\mathrm{Tr}\left(  e^{t\Delta_{1}}-e^{t\Delta_{2}}\right)
\,,\label{int-proper-time}%
\end{equation}
one can identify the divergence of integral (\ref{int-proper-time}) at small
$t$ with the contribution of cusps into the ultraviolet divergences of the
determinant $\mathrm{Det}\,\Delta_{C}$ of the Laplace operator.

In the LSW model \cite{Luescher-80}, \cite{Luescher-81}, the determinant of
the two-dimensional Laplace operator $\Delta_{C}$ with Dirichlet boundary
conditions on the contour $C$ plays the central role: flat Wilson loops $W(C)$
of the $D$ dimensional gauge theory are approximated in the LSW model by%
\begin{equation}
W_{\mathrm{LSW}}(C)=e^{-KS(C)}\left(  \mathrm{Det}\,\Delta_{C}\right)
^{-(D-2)/2}\,. \label{W-C-det}%
\end{equation}
Combining rigorous result (\ref{sp-exp-expansion}) and model expression
(\ref{W-C-det}), one obtains the interpretation of the function $\Gamma
_{\mathrm{cusp}}^{\mathrm{str}}(\theta)$ as a cusp anomalous dimension in the
effective string model for Wilson loops.

Now one arrives at the conjecture that the convexity of the $\Gamma
_{\mathrm{cusp}}^{\mathrm{str}}(\theta)$ (\ref{Gamma-1-convex-string}) is not
accidental. It may happen that the inequalities derived in this paper in the
framework of the gauge theory also hold in the LSW model. In particular, one
can show that Bachas inequality (\ref{Bachas-ineq}) (and some of its
generalizations) really holds in the model:%
\begin{equation}
\left[  W_{\mathrm{LSW}}\left(  T,\frac{R_{1}+R_{2}}{2}\right)  \right]
^{2}\leq W_{\mathrm{LSW}}\left(  T,R_{1}\right)  W_{\mathrm{LSW}}\left(
T,R_{2}\right)  \,.
\end{equation}
According to Eq.~(\ref{W-C-det}) this inequality is equivalent to the
following inequality for the determinant of the Laplace operator $\Delta(T,R)$
with Dirichlet boundary conditions on the rectangle $T\times R$:%
\begin{equation}
\left[  \mathrm{Det}\,\Delta\left(  T,\frac{R_{1}+R_{2}}{2}\right)  \right]
^{2}\geq\left[  \mathrm{Det}\,\Delta\left(  T,R_{1}\right)  \right]  \left[
\mathrm{Det}\,\Delta\left(  T,R_{2}\right)  \right]  \,.
\end{equation}

In the above argument, the LSW model played an important heuristic role.
However, one has to separate the interesting physical question about the
relevance of the LSW model for large Wilson loops in gauge theories
\cite{Alvarez-81,LW-02,Greensite-03,Caselle-05,Meyer-06} from the formal
mathematical properties\footnote{Certainly the LSW model has nothing to do
with the true small distance physics controlling the cusp behavior in the
gauge theory. However, the formal mathematical extrapolation of the LSW model
expressions to nonphysical contours (like the above discussion of the
connection between the convexity of the cusp anomalous dimension and possible
inequalities in the LSW model) can be used as a heuristic way to approach the
properties of the determinants of Laplacians. From the rigorous point of view,
these conjectures about the determinants should be checked mathematically
without any reference to the LSW model.} of the determinants of Laplacians.
Now we can formulate the mathematical side of the problem in a precise form
without appealing to the LSW model: Do inequalities \emph{derived in this
paper in the framework of the gauge theory} also hold if we replace all Wilson
loops $W(C)$ by the determinants of Laplace operators $\left(  \mathrm{Det}%
\,\Delta_{C}\right)  ^{-\nu}$ with $\nu>0$? A partial positive answer to this
interesting question will be given in Sec.~\ref{String-section} for some
limited class of inequalities.

\subsection{Structure of the paper}

In Sec.~\ref{Trapezium-parallelogram-section} inequality
(\ref{parall-trap-ineq}) is derived (Sec.~\ref{Derivation-t-p-subsection}) and
its compatibility with the area law is checked (Sec.~\ref{area-law-subsection}).

Sec.~\ref{Renormalization-cusps-section} is devoted to the renormalization of
inequalities. In Sec.~\ref{UV-divergence-subsection} the basic facts about the
renormalization of Wilson loops are briefly described. In Sec.
\ref{triangle-limit-subsection} we analyze the limit when the trapezium is
contracted into a triangle in inequality (\ref{parall-trap-ineq}). In this
singular limit inequalities become trivial but consistent. We show that this
consistency is guaranteed by the convexity of the cusp anomalous dimension
(Sec.~\ref{convexity-AD-subsection}). Generalizing the example of the singular
trapezium-to-triangle limit, we describe in Sec.~\ref{survival-subsection} two
classes of inequalities \emph{quadratic} in Wilson loops: one group of
inequalities survives in a non-trivial form after the renormalization whereas
the second group reduces to trivial but correct inequalities.

In Sec.~\ref{More-general-section} we show how more general \emph{polynomial}
inequalities can derived. In Sec.~\ref{generalization-derivation-subsection}
we derive these inequalities in the form of positivity constraints on
determinants of matrices of different Wilson loops. As an example, we consider
a generalization of the trapezium-parallelogram inequality in Sec.
\ref{general-example-subsection}. In Sec.~\ref{general-area-law-subsection} we
check the compatibility of inequalities with the area law. The renormalization
properties of these general inequalities are described in Sec.
\ref{renormalization-general-subsection} and proved in Appendix
\ref{Appendix-section}. In Sec.~\ref{Further-generalizations-subsection} we
comment on specific features of inequalities for non-flat Wilson loops. In
Sec.~\ref{Positivity-spectral-subsection} we discuss the connection between
the spectral representations for Wilson loops and inequalities.

Sec.~\ref{String-section} contains several interesting observations concerning
the LSW model for Wilson loops \cite{Luescher-80,Luescher-81}. In this model
cusp singularities of polygonal contours are described by an anomalous
dimension which is different from the gauge theories but still has convexity
property (Sec.~\ref{Cusp-string-subsection}). The rectangular loops computed
in this string model (Sec.~\ref{Rectangular-string-subsection}) obey
inequalities derived for Wilson loops in the context of the gauge theory (Sec.
\ref{Spectral-string-section}).

\section{Trapezium-parallelogram inequality}

\label{Trapezium-parallelogram-section}

\setcounter{equation}{0} 

\subsection{Derivation}

\label{Derivation-t-p-subsection}

Our derivation of inequalities is a trivial modification of the original
method used by Bachas in Ref. \cite{Bachas-86}. Let us consider two open paths
$B_{1}$ and $B_{2}$ with common end points $x,y$ at zero Euclidean time:%
\begin{equation}
x_{4}=y_{4}=0\,.
\end{equation}
We assume that both paths are placed in the semi-space with negative Euclidean
times and only the end points reach the zero-time hyperplane as shown in Fig.
\ref{fig-2} (c). We use notation $U_{ab}(B_{k})$ for the operator Wilson lines
$U_{ab}(B_{k})$ taken along the lines $B_{k}$. Here $a,b$ are gauge indices
associated with the open ends of the line $B_{k}$.

Then the positivity of the squared norm of the linear combination%
\begin{equation}
\left\|  \sum\limits_{ab}\left[  k_{1}^{ab}U_{ab}(B_{1})|0\rangle+k_{2}%
^{ab}U_{ab}(B_{2})\right]  |0\rangle\right\|  ^{2}\geq0\label{L-B-ineq}%
\end{equation}
with arbitrary coefficients $k_{1}^{ab}$, $k_{2}^{ab}$ leads to the Cauchy
inequality%
\begin{equation}
\left|  w_{12}\right|  ^{2}\leq w_{11}w_{22}\label{w-Cauchy-ineq}%
\end{equation}
where%
\begin{equation}
w_{kl}=\sum\limits_{ab}\langle0|\left[  U_{ab}(B_{l})\right]  ^{+}U_{ab}%
(B_{k})|0\rangle\,.
\end{equation}
Obviously%
\begin{equation}
W(C_{kl})=\frac{1}{N}w_{kl}%
\end{equation}
is nothing else but the Wilson line (\ref{W-C-general}) corresponding to the
closed contour made of $B_{k}$ with the time-reflected path $B_{l}^{T}$:%
\begin{equation}
C_{kl}=B_{k}\cup B_{l}^{T}\,.\label{C-kl-via-B}%
\end{equation}
Now inequality (\ref{w-Cauchy-ineq}) takes the form%
\begin{equation}
\left|  W(C_{12})\right|  ^{2}\leq W(C_{11})W(C_{22}%
)\,.\label{W-C-ineq-general}%
\end{equation}
Choosing the open paths $B_{k}$ as shown in Fig.~\ref{fig-2} (c), we obtain%
\begin{equation}
W(C_{11})=W(C_{22})=W_{\mathrm{trapezium}}(a,b,h)\,,
\end{equation}%
\begin{equation}
W(C_{12})=W_{\mathrm{parallelogram}}\left(  \frac{a+b}{2},h\right)  \,.
\end{equation}
The parallelogram-trapezium inequality (\ref{parall-trap-ineq}) immediately
follows from inequality (\ref{W-C-ineq-general}).

In the above derivation we could choose arbitrary lines $B_{k}$ (with common
end points on the zero time hyperplane and all other points having negative
time). Therefore inequality (\ref{W-C-ineq-general}) is valid for arbitrary
composite contours $C_{kl}$ made from open lines $B_{k}$ according to
(\ref{C-kl-via-B}).

\begin{figure}
[ptb]
\begin{center}
\includegraphics[
height=1.2073in,
width=1.5471in
]%
{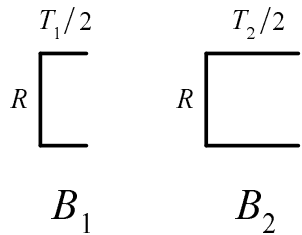}%
\caption{Open paths used for the derivation of Bachas inequality
(\ref{Bachas-ineq}).}%
\label{fig-3}%
\end{center}
\end{figure}

Taking inequality (\ref{W-C-ineq-general}) with open lines $B_{1}$, $B_{2}$
shown in Fig.~\ref{fig-3}, one obtains the inequality for rectangular loops%
\begin{equation}
\left[  W\left(  \frac{T_{1}+T_{2}}{2},R\right)  \right]  ^{2}\leq W\left(
T_{1},R\right)  W\left(  T_{2},R\right)  \,,
\end{equation}
which due to the symmetry%
\begin{equation}
W\left(  T,R\right)  =W\left(  R,T\right)  \,,\label{T-R-symmetry-W}%
\end{equation}
reproduces Bachas inequality (\ref{Bachas-ineq}).%

\subsection{Compatibility with the area law}

\label{area-law-subsection}

Inequality (\ref{parall-trap-ineq}) was derived using only the positivity of
the norm for physical states. Therefore this inequality is general. It holds
both in gauge theories with confinement and in theories without confinement.
Still we can check the compatibility of this inequality with the area law. The
standard dogma of the area law for the large-size Wilson loops in a confining
non-Abelian pure gauge theory reads%
\begin{equation}
W(C)=A(C)\,\exp\left[  -a_{0}\,L(C)-KS(C)\right]  \,.\label{area-law-general}%
\end{equation}
Here $L(C)$, $S(C)$ are the perimeter and the surface area of a large flat
contour $C$, whereas the prefactor $A(C)$ may have only a nonexponential
behavior. Coefficients $a_{0}$, $K$ are the same as in the large distance
expansion of the static potential (\ref{V-large-R}).

Now we can turn to the question of the compatibility of the area law
(\ref{area-law-general}) with inequality (\ref{parall-trap-ineq}). Note that
the parallelogram and the trapezium have the same perimeter and area in Eq.
(\ref{parall-trap-ineq}). Therefore inserting (\ref{area-law-general}) into
inequality (\ref{parall-trap-ineq}), we find%
\begin{equation}
A_{\mathrm{parallelogram}}\left(  \frac{a+b}{2},h\right)  \,\leq
A_{\mathrm{trapezium}}(a,b,h)\,. \label{A-parallelogram-A-trapezium}%
\end{equation}
Thus the main result is that the exponential area and perimeter growth can be
factored out from the inequality.

\section{Renormalization and cusp singularities}

\label{Renormalization-cusps-section}

\setcounter{equation}{0} 

\subsection{Renormalization}

\label{UV-divergence-subsection}

Some care must be taken about the ultraviolet divergences and their
renormalization. As is well known
\cite{Polyakov-79,DV-80,Arefeva-80,BNS-81,BGSN-82}, the renormalization of
closed Wilson loops involves two factors:

1) perimeter renormalization constant $Z_{\mathrm{perimeter}}(\Lambda,L)$
depending on the length of the contour $L$ and on the ultraviolet cutoff
$\Lambda$ with the exponential dependence on $L$:%
\begin{equation}
Z_{\mathrm{per.}}(\Lambda,L_{1}+L_{2})=Z_{\mathrm{per.}}(\Lambda
,L_{1})Z_{\mathrm{per.}}(\Lambda,L_{2})\,, \label{Z-per-mult}%
\end{equation}

2) cusp renormalization constant factorizable in contributions of separate
cusps with angles $\gamma_{k}$ (Fig.~\ref{fig-1})%
\begin{equation}
\prod_{k}Z_{\mathrm{cusp}}(\Lambda,\gamma_{k})\,.
\end{equation}

The renormalization of $W(C)$ is described by the equation
\cite{Polyakov-79,DV-80,Arefeva-80,BNS-81,BGSN-82}%
\begin{equation}
W^{\mathrm{ren}}(C)=\lim_{\Lambda\rightarrow\infty}Z(\Lambda
,C)W^{\mathrm{nonren}}(C,\Lambda) \label{W-renormalization}%
\end{equation}
where%
\begin{equation}
Z(\Lambda,C)=Z_{\mathrm{per.}}(\Lambda,L(C))\prod_{C\,\mathrm{cusps}%
}Z_{\mathrm{cusp}}(\Lambda,\gamma)\,. \label{Z-per-cusp}%
\end{equation}
Now we see that%
\begin{equation}
Z_{\mathrm{trapezium}}(\Lambda)=Z_{\mathrm{parallelogram}}(\Lambda)
\end{equation}
because our parallelogram and trapezium have the same perimeter and the same
set of cusp angles.

Therefore renormalization does not change inequality (\ref{parall-trap-ineq}).

\subsection{Singularity of the triangle limit}

\label{triangle-limit-subsection}%

\begin{figure}
[ptb]
\begin{center}
\includegraphics[
height=1.1381in,
width=3.6365in
]%
{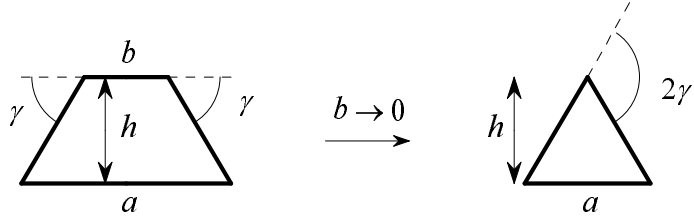}%
\caption{Limit $b\rightarrow0$ converting a trapezium into a triangle.}%
\label{fig-4}%
\end{center}
\end{figure}

Now let us give an example of an inequality which becomes trivial after
renormalization. At first sight, one could take the limit $b\rightarrow0$ in
inequality (\ref{parall-trap-ineq}) as shown in Fig.~\ref{fig-4} and obtain
the inequality for the Wilson loop corresponding to the triangle with the base
$a$ and height $h$:%
\begin{equation}
\left|  W_{\mathrm{parallelogram}}^{\mathrm{nonren}}\left(  \frac{a}%
{2},h;\Lambda\right)  \right|  \leq W_{\mathrm{triangle}}^{\mathrm{nonren}%
}(a,h;\Lambda)\,.\label{triangle-ineq-nonren}%
\end{equation}
This inequality is derived for regularized and nonrenormalized Wilson loops.
The renormalization of the Wilson loops is described by equations
(\ref{W-renormalization}), (\ref{Z-per-cusp}):%
\begin{align}
W_{\mathrm{triangle}}^{\mathrm{ren}}(a,h) &  =\lim_{\Lambda\rightarrow\infty
}Z_{\mathrm{per.}}(\Lambda,L_{\mathrm{triangle}})\nonumber\\
&  \times\left[  \prod_{k=1}^{3}Z_{\mathrm{cusp}}(\Lambda,\gamma
_{k}^{\mathrm{triangle}})\right]  W_{\mathrm{triangle}}^{\mathrm{nonren}%
}(a,h;\Lambda)\\
W_{\mathrm{parallelogram}}^{\mathrm{ren}}\left(  \frac{a}{2},h\right)   &
=\lim_{\Lambda\rightarrow\infty}Z_{\mathrm{per.}}(\Lambda
,L_{\mathrm{parallelogram}})\nonumber\\
&  \times\left[  \prod_{k=1}^{4}Z_{\mathrm{cusp}}(\Lambda,\gamma
_{k}^{\mathrm{parallelogram}})\right]  W_{\mathrm{parallelogram}%
}^{\mathrm{nonren}}\left(  \frac{a}{2},h;\Lambda\right)  \,.
\end{align}
Note that the triangle and the parallelogram have the same perimeter%
\begin{equation}
L_{\mathrm{triangle}}=L_{\mathrm{parallelogram}}%
\end{equation}
so that the contribution of the perimeter renormalization cancels in the
renormalized version of inequality (\ref{triangle-ineq-nonren}). However, the
set of cusp angles is different:%
\begin{equation}
\frac{\prod_{k=1}^{4}Z_{\mathrm{cusp}}(\Lambda,\gamma_{k}%
^{\mathrm{parallelogram}})}{\prod_{k=1}^{3}Z_{\mathrm{cusp}}(\Lambda
,\gamma_{k}^{\mathrm{triangle}})}=\frac{\left[  Z_{\mathrm{cusp}}%
(\Lambda,\gamma)\right]  ^{2}}{Z_{\mathrm{cusp}}(\Lambda,2\gamma)}\,.
\end{equation}
Here we noticed that the mismatch between cusp angles involves two angles
$\gamma$ of the parallelogram and one angle of the triangle $2\gamma$ which is
twice larger (see Fig.~\ref{fig-4}).

Now the renormalized version of inequality (\ref{triangle-ineq-nonren})
becomes%
\begin{equation}
\left|  W_{\mathrm{parallelogram}}^{\mathrm{ren}}\left(  \frac{a}{2},h\right)
\right|  \leq W_{\mathrm{triangle}}^{\mathrm{ren}}(a,h)\lim_{\Lambda
\rightarrow\infty}\frac{\left[  Z_{\mathrm{cusp}}(\Lambda,\gamma)\right]
^{2}}{Z_{\mathrm{cusp}}(\Lambda,2\gamma)}\,.\label{triangle-ineq-non}%
\end{equation}
The ratio of renormalization constants $Z_{\mathrm{cusp}}(\Lambda,\gamma)$ has
a singular behavior in the limit $\Lambda\rightarrow\infty$ and has a
nontrivial dependence on $\gamma$. In principle, two cases could be expected%
\begin{equation}
\lim_{\Lambda\rightarrow\infty}\frac{\left[  Z_{\mathrm{cusp}}(\Lambda
,\gamma)\right]  ^{2}}{Z_{\mathrm{cusp}}(\Lambda,2\gamma)}=\left\{
\begin{array}
[c]{c}%
\infty,\\
0.
\end{array}
\right.  \label{Z-ratio-two-cases}%
\end{equation}
In the first case of the infinite limit, inequality (\ref{triangle-ineq-non})
would be reduced to the triviality%
\begin{equation}
W_{\mathrm{parallelogram}}^{\mathrm{ren}}\left(  \frac{a}{2},h\right)
\leq\infty\,.\label{ineq-trivial}%
\end{equation}
In the second case of the zero limit in Eq.~(\ref{Z-ratio-two-cases}) we would
obtain a rather improbable result:%
\begin{equation}
W_{\mathrm{parallelogram}}^{\mathrm{ren}}\left(  \frac{a}{2},h\right)
=0\,.\label{triangle-second-case}%
\end{equation}
From this point of view one would expect that the first case in
(\ref{Z-ratio-two-cases})%
\begin{equation}
\lim_{\Lambda\rightarrow\infty}\frac{\left[  Z_{\mathrm{cusp}}(\Lambda
,\gamma)\right]  ^{2}}{Z_{\mathrm{cusp}}(\Lambda,2\gamma)}=\infty
\label{Z-phi-lim}%
\end{equation}
is more natural.

In asymptotically free theories this limit can be studied perturbatively. In
the next section we will show that in asymptotically free non-Abelian
SU($N_{c}$) gauge theory we have%
\begin{equation}
\lim_{\Lambda\rightarrow\infty}\frac{\left[  Z_{\mathrm{cusp}}\left(
\Lambda,\gamma+\gamma^{\prime}\right)  \right]  ^{2}}{Z_{\mathrm{cusp}}\left(
\Lambda,2\gamma\right)  Z_{\mathrm{cusp}}\left(  \Lambda,2\gamma^{\prime
}\right)  }=\left\{
\begin{array}
[c]{cc}%
\infty & \mathrm{if}\quad\,\gamma\neq\gamma^{\prime}\,,\\
1 & \mathrm{if}\quad\gamma=\gamma^{\prime}\,.
\end{array}
\right.  \label{Z-convexity}%
\end{equation}
in the interval
\begin{equation}
-\frac{\pi}{2}\leq\gamma,\gamma^{\prime}\leq\frac{\pi}{2}\,.
\end{equation}
Taking $\gamma^{\prime}=0$ in this relation (which means no cusp so that
$Z_{\mathrm{cusp}}\left(  \Lambda,2\gamma^{\prime}\right)  =1$), we obtain
divergence (\ref{Z-phi-lim}).

We see that the limit $b\rightarrow0$ in the renormalized version of
trapezium-parallelogram inequality (\ref{triangle-ineq-nonren}) leads to
triviality (\ref{ineq-trivial}). Although our attempt to derive an inequality
for renormalized triangular Wilson loops has failed, the fact that one avoids
the pathological case (\ref{triangle-second-case}) demonstrates the
compatibility of inequality (\ref{parall-trap-ineq}) with the properties of
the cusp anomalous dimension controlling the limit (\ref{Z-phi-lim}). These
properties will be considered in the next section.

\subsection{Convexity of the cusp anomalous dimension}

\label{convexity-AD-subsection}

Let us compute the limit (\ref{Z-convexity}). In the asymptotically free case
(e.g. QCD with a limited number of quark flavors $N_{f}<11N_{c}/2$) this limit
is determined by the leading order of the cusp anomalous dimension
(\ref{Gamma-1-cusp})%

\begin{align}
\Gamma_{\mathrm{cusp}}^{(1)}(g,\gamma) &  =\frac{g^{2}}{4\pi^{2}}\Gamma
(\gamma)\,,\\
\Gamma(\gamma) &  =-4C_{F}\left(  1-\gamma\cot\gamma\right)  \,,\\
C_{F} &  =\frac{N_{c}^{2}-1}{2N_{c}}\,
\end{align}
where $\gamma$ is the cusp angle dual to the inside-facing angle $\pi-\gamma$
as shown in Fig.~\ref{fig-1}. Starting from the trivial inequalities%
\begin{gather}
0<\gamma<\tan\gamma\quad\left(  0<\gamma<\frac{\pi}{2}\right)  \,,\\
\cot\gamma\leq0\quad\left(  \frac{\pi}{2}\leq\gamma<\pi\right)  \,,
\end{gather}
one finds%
\begin{equation}
1-\gamma\cot\gamma>0\,.
\end{equation}
for $0<\gamma<\pi$. Since function $1-\gamma\cot\gamma$ is even, we can extend
the validity interval to%
\begin{equation}
-\pi<\gamma<\pi\label{gamma-interval}%
\end{equation}
except for the point $\gamma=0$. Now we can prove the positivity of the second
derivative:%
\begin{equation}
\frac{d^{2}}{d^{2}\gamma}\left(  1-\gamma\cot\gamma\right)  =\left(
1-\gamma\cot\gamma\right)  \frac{2}{\sin^{2}\gamma}>0\,.
\end{equation}
This shows that%
\begin{align}
\Gamma(\gamma) &  <0\,\quad(-\pi<\gamma<\pi,\,\quad\gamma\neq0)\,,\\
\frac{d^{2}}{d^{2}\gamma}\Gamma(\gamma) &  <0\quad(-\pi<\gamma<\pi)\,.
\end{align}
Hence function $\Gamma(\gamma)$ is convex%
\begin{equation}
\Gamma(\gamma+\gamma^{\prime})>\frac{1}{2}\left[  \Gamma(2\gamma
)+\Gamma(2\gamma^{\prime})\right]  \quad\left(  -\pi/2<\gamma\neq
\gamma^{\prime}<\pi/2\right)  \,.\label{Gamma-convex}%
\end{equation}
This is nothing else but convexity property (\ref{Gamma-1-convex}) which was
already mentioned in the introduction.

The asymptotic behavior of the renormalization constant at large cutoff
$\Lambda$ is%
\begin{equation}
Z_{\mathrm{cusp}}(\Lambda,\gamma)\overset{\Lambda\rightarrow\infty}{\sim
}\left[  g(\Lambda)\right]  ^{-\Gamma(\gamma)/\beta_{1}}
\label{Z-cusp-large-Lambda}%
\end{equation}
where $\beta_{1}$ is the first coefficient of the beta function%
\begin{equation}
\beta_{1}=\frac{11}{3}N_{c}-\frac{2}{3}N_{f}\,.
\end{equation}

In the limit $\Lambda\rightarrow\infty$, using (\ref{Gamma-convex}) and
$g(\Lambda)\rightarrow0$, we find for $-\pi/2<\gamma\neq\gamma^{\prime}<\pi/2$%
\begin{equation}
\frac{\left[  Z_{\mathrm{cusp}}\left(  \Lambda,\gamma+\gamma^{\prime}\right)
\right]  ^{2}}{Z_{\mathrm{cusp}}\left(  \Lambda,2\gamma\right)
Z_{\mathrm{cusp}}\left(  \Lambda,2\gamma^{\prime}\right)  }\sim\left[
g^{2}(\Lambda)\right]  ^{-\left[  2\Gamma(\gamma+\gamma^{\prime}%
)-\Gamma(2\gamma)-\Gamma(2\gamma^{\prime})\right]  /\beta_{1}}\rightarrow
\infty\,.
\end{equation}
Thus the limit (\ref{Z-convexity}) is computed (the case $\gamma
=\gamma^{\prime}$ trivial). In particular, this completes the derivation of
relation (\ref{Z-phi-lim}) which was used for the analysis of the
singularities accompanying the transformation of a trapezium into a triangle.

\subsection{Which inequalities remain nontrivial in the continuum limit}

\label{survival-subsection}

The above derivation of inequality (\ref{W-C-ineq-general}) can be generalized
for arbitrary open paths $B_{1}$, $B_{2}$ but one should take care about the
renormalization stability of inequalities. The nonrenormalized inequality%
\begin{equation}
\left[  W^{\mathrm{nonren}}(C_{12},\Lambda)\right]  ^{2}\leq
W^{\mathrm{nonren}}(C_{11},\Lambda)W^{\mathrm{nonren}}(C_{22},\Lambda)
\end{equation}
is modified by the renormalization (\ref{W-renormalization}):%
\begin{equation}
\left[  W^{\mathrm{ren}}(C_{12})\right]  ^{2}\leq W^{\mathrm{ren}}%
(C_{11})W^{\mathrm{ren}}(C_{22})\,\lim_{\Lambda\rightarrow\infty}\frac{\left[
Z(\Lambda,C_{12})\right]  ^{2}}{Z(\Lambda,C_{11})Z(\Lambda,C_{22}%
)}\,.\label{ineq-renormalization-1}%
\end{equation}

In the simplest case when the two open paths $B_{1}$ and $B_{2}$ reach the end
points with the tangent lines directed along the Euclidean time axis [e.g. the
cases shown in Figs. \ref{fig-2}(c) and \ref{fig-3}] this stability is trivial:

1) The perimeters of contours $C_{kl}$ in (\ref{W-C-ineq-general}) obey
condition%
\begin{equation}
2L(C_{12})=L(C_{11})+L(C_{22})=2L(B_{1})+2L(B_{2})\,.
\end{equation}
Inserting this into (\ref{Z-per-mult}), we obtain%
\begin{equation}
Z_{\mathrm{per.}}(\Lambda,L(C_{11}))Z_{\mathrm{per.}}(\Lambda,L(C_{22}%
))\,=\left[  Z_{\mathrm{per.}}(\Lambda,L(C_{12}))\right]  ^{2}.
\label{Z-per-mult-2}%
\end{equation}

2) The combined set of cusp angles of contours $C_{11}$ and $C_{22}$ coincides
with the double set of cusp angles of the contour $C_{12}$. Therefore%
\begin{equation}
\left[  \prod_{C_{11}\,\mathrm{cusps}}Z_{\mathrm{cusp}}(\Lambda,\gamma
)\right]  \left[  \prod_{C_{22}\,\mathrm{cusps}}Z_{\mathrm{cusp}}%
(\Lambda,\gamma)\right]  =\left[  \prod_{C_{12}\,\mathrm{cusps}}%
Z_{\mathrm{cusp}}(\Lambda,\gamma)\right]  ^{2}\,. \label{Z-cusp-mult-2}%
\end{equation}
Inserting Eqs. (\ref{Z-per-mult-2}), (\ref{Z-cusp-mult-2}) into Eq.
(\ref{Z-per-cusp}), we find that the renormalization constants for contours
$C_{kl}$ obey relation%
\begin{equation}
Z(\Lambda,C_{11})Z(\Lambda,C_{22})=\left[  Z(\Lambda,C_{12})\right]  ^{2}%
\end{equation}
so that renormalization (\ref{ineq-renormalization-1}) does not change
inequality (\ref{W-C-ineq-general}).%

\begin{figure}
[ptb]
\begin{center}
\includegraphics[
height=5.2529in,
width=4.5333in
]%
{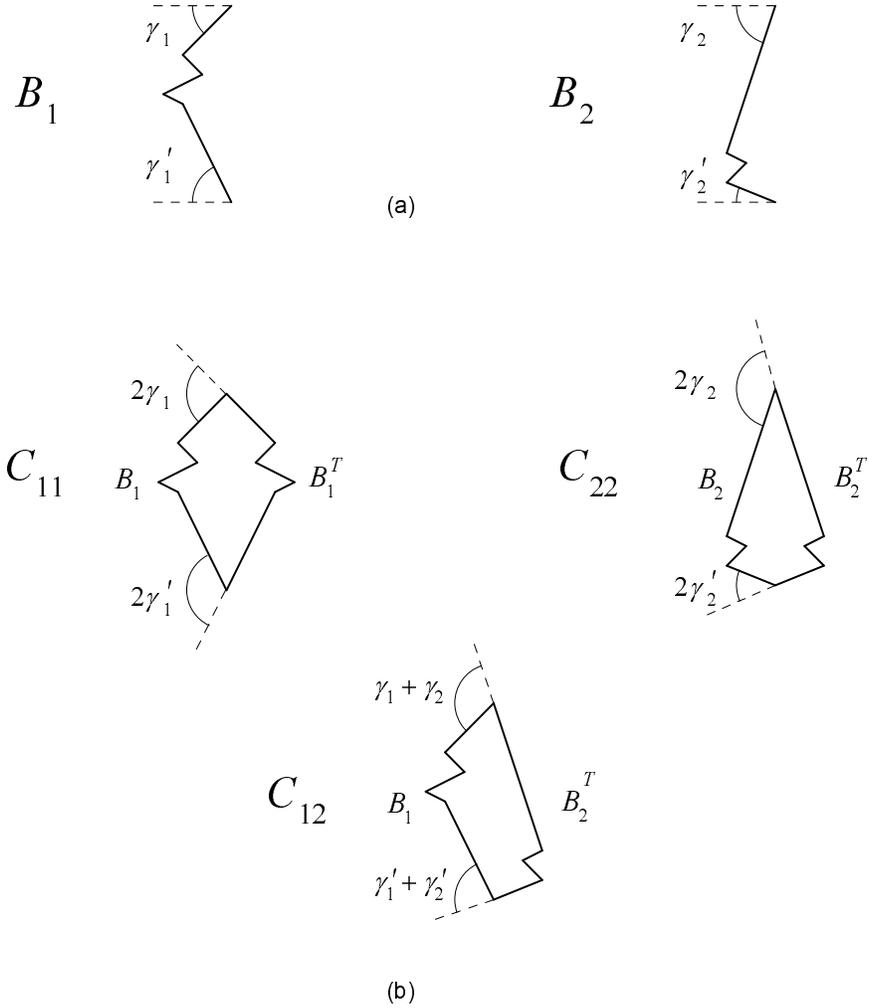}%
\caption{(a) Open lines $B_{k}$ and their end point cusp angles $\gamma
_{k},\gamma_{k}^{\prime}$ with respect to the horizontal Euclidean time axis.
Note that angles $\gamma_{k},\gamma_{k}^{\prime}$ can be negative. (b)
Corresponding closed contours $C_{kl}$ with the cusp angles $\gamma_{k}%
+\gamma_{l}$ and $\gamma_{k}^{\prime}+\gamma_{l}^{\prime}$.}%
\label{fig-5}%
\end{center}
\end{figure}

The situation becomes more interesting if we allow for lines $B_{1}$ and
$B_{2}$ which have arbitrary directions of tangent vectors at the end points.
In this case%
\begin{equation}
\sum\limits_{C_{kl}\,\mathrm{cusps}}\Gamma(\gamma)=\sum
\limits_{\substack{\mathrm{internal}\\\,B_{k}\,\mathrm{cusps}}}\Gamma
(\gamma)+\sum\limits_{\substack{\mathrm{internal}\\\,B_{l}\,\mathrm{cusps}%
}}\Gamma(\gamma)+\Gamma(\gamma_{k}+\gamma_{l})+\Gamma(\gamma_{k}^{\prime
}+\gamma_{l}^{\prime})\,. \label{C-kl-cusps-general}%
\end{equation}
where $\gamma_{k}$ and $\gamma_{k}^{\prime}$ are ``end-point cusp angles'' of
open lines $B_{k}$ counted from the Euclidean time axis as shown in Fig.
\ref{fig-5} (a). These end-point cusp angles of $B_{k}$ and $B_{l}$ generate
additional cusp angles $\gamma_{k}+\gamma_{l}$ and $\gamma_{k}^{\prime}%
+\gamma_{l}^{\prime}$ of the closed contour $C_{kl}$ (see Fig.~\ref{fig-5}
(b)). Note that angles $\gamma_{k},\gamma_{k}^{\prime}$ can be positive or
negative:%
\begin{equation}
-\frac{\pi}{2}<\gamma_{k},\gamma_{k}^{\prime}<\frac{\pi}{2}%
\end{equation}
The choice of the positive direction for counting angles $\gamma_{k}$ does not
matter ($\Gamma(\gamma)$ is an even) but this choice should be the same for
all lines $B_{k}$.

Using Eq.~(\ref{C-kl-cusps-general}), we see that the internal cusp angles of
lines $B_{k}$ cancel in the combination%
\begin{align}
&  \sum\limits_{C_{11}\,\mathrm{cusps}}\Gamma(\gamma)+\sum\limits_{C_{22}%
\,\mathrm{cusps}}\Gamma(\gamma)-2\sum\limits_{C_{12}\,\mathrm{cusps}}%
\Gamma(\gamma)\nonumber\\
&  =\left[  \Gamma(2\gamma_{1})+\Gamma(2\gamma_{1}^{\prime})\right]  +\left[
\Gamma(2\gamma_{2})+\Gamma(2\gamma_{2}^{\prime})\right]  -2\left[
\Gamma(\gamma_{1}+\gamma_{2})+\Gamma(\gamma_{1}^{\prime}+\gamma_{2}^{\prime
})\right]  \nonumber\\
&  =\left[  \Gamma(2\gamma_{1})+\Gamma(2\gamma_{2})-2\Gamma(\gamma_{1}%
+\gamma_{2})\right]  +\left[  \Gamma(2\gamma_{1}^{\prime})+\Gamma(2\gamma
_{2}^{\prime})-2\Gamma(\gamma_{1}^{\prime}+\gamma_{2})\right]  \,.
\end{align}
On the RHS we have two terms which are almost strictly negative according to
inequality (\ref{Gamma-convex}):%
\begin{align}
\Gamma(2\gamma_{1})+\Gamma(2\gamma_{2})-2\Gamma(\gamma_{1}+\gamma_{2}) &
<0\quad\mathrm{if}\quad\quad-\pi/2<\gamma_{1}\neq\gamma_{2}<\pi
/2\,,\label{Gamma-convexity-1}\\
\Gamma(2\gamma_{1}^{\prime})+\Gamma(2\gamma_{2}^{\prime})-2\Gamma(\gamma
_{1}^{\prime}+\gamma_{2}^{\prime}) &  <0\quad\mathrm{if}\quad\quad
-\pi/2<\gamma_{1}^{\prime}\neq\gamma_{2}^{\prime}<\pi
/2\,.\label{Gamma-convexity-2}%
\end{align}
Therefore%
\begin{equation}
\sum\limits_{C_{11}\,\mathrm{cusps}}\Gamma(\gamma)+\sum\limits_{C_{22}%
\,\mathrm{cusps}}\Gamma(\gamma)-2\sum\limits_{C_{12}\,\mathrm{cusps}}%
\Gamma(\gamma)\left\{
\begin{array}
[c]{l}%
<0\quad\mathrm{if}\,\,\gamma_{1}\neq\gamma_{2}\,\,\mathrm{or}\,\,\gamma
_{1}^{\prime}\neq\gamma_{2}^{\prime}\,,\\
=0\quad\mathrm{if}\,\,\gamma_{1}=\gamma_{2}\,\,\mathrm{and}\,\,\gamma
_{1}^{\prime}=\gamma_{2}^{\prime}\,.
\end{array}
\right.
\end{equation}
Combining this with Eq.~(\ref{Z-cusp-large-Lambda}), we find%
\begin{align}
&  \lim_{\Lambda\rightarrow\infty}\frac{\left[  Z(\Lambda,C_{12})\right]
^{2}}{Z(\Lambda,C_{11})Z(\Lambda,C_{22})}\nonumber\\
&  =\lim_{\Lambda\rightarrow\infty}\exp\left\{  \frac{\ln g^{2}}{\beta_{1}%
}\left[  \sum\limits_{C_{11}\,\mathrm{cusps}}\Gamma(\gamma)+\sum
\limits_{C_{22}\,\mathrm{cusps}}\Gamma(\gamma)-2\sum\limits_{C_{12}%
\,\mathrm{cusps}}\Gamma(\gamma)\right]  \right\}  \nonumber\\
&  =\left\{
\begin{array}
[c]{l}%
\infty\quad\mathrm{if}\quad\gamma_{1}\neq\gamma_{2}\quad\mathrm{or}\quad
\gamma_{1}^{\prime}\neq\gamma_{2}^{\prime}\,,\\
1\quad\mathrm{if}\quad\gamma_{1}=\gamma_{2}\quad\mathrm{and}\quad\gamma
_{1}^{\prime}=\gamma_{2}^{\prime}\,.
\end{array}
\right.  \label{lim-Z-ratio}%
\end{align}
Inserting this into inequality (\ref{ineq-renormalization-1}), we see that a
nontrivial result comes after the renormalization only in the case%
\begin{equation}
\gamma_{1}=\gamma_{2},\quad\gamma_{1}^{\prime}=\gamma_{2}^{\prime
}\,.\label{cusp-balance}%
\end{equation}
Otherwise after renormalization we arrive at the triviality%
\begin{equation}
\left|  W(C_{12})\right|  \leq\infty\,.\label{W-12-triviality}%
\end{equation}
We conclude that inequality (\ref{W-C-ineq-general}) is not changed by the
renormalization only in the case when the combined set of cusp angles of
contours $C_{11}$ and $C_{22}$ coincides with the double set of the cusp
angles of contour $C_{12}$. If this condition is violated then we arrive at
the \emph{correct but useless} result (\ref{W-12-triviality}). Whatever
trivial this result is, it is important for us that the ratio of
renormalization constants (\ref{lim-Z-ratio}) diverges to infinity and does
not go to zero as it could be in principle. If we had a zero limit in
(\ref{lim-Z-ratio}) then this would correspond to a rather improbable
situation with many vanishing Wilson loops. Now one has to remember that the
divergence of the ratio (\ref{lim-Z-ratio}) and the protection from the zero
limit originates from the convexity property of the cusp anomalous dimension
which was used in Eqs. (\ref{Gamma-convexity-1}), (\ref{Gamma-convexity-2}).
Thus we see that the convexity of the cusp anomalous dimension ``protects''
the consistency of the theory when one tries to take singular limits in
inequalities for Wilson loops.

\section{More general inequalities}

\setcounter{equation}{0} 

\label{More-general-section}

\subsection{Derivation}

\label{generalization-derivation-subsection}

We can generalize inequality (\ref{L-B-ineq}) by taking an arbitrary amount of
open paths $B_{k}$ with common end points:%
\begin{equation}
\left\|  \sum\limits_{m=1}^{M}\sum\limits_{ab}\left[  k_{m}^{ab}U_{ab}%
(B_{m})\right]  |0\rangle\right\|  ^{2}\geq0\,.
\end{equation}
This means that matrix%
\begin{equation}
W(C_{kl})=\frac{1}{N}\sum\limits_{ab}\langle0|\left[  U_{ab}(B_{l})\right]
^{+}U_{ab}(B_{k})|0\rangle
\end{equation}
is positive definite. In particular, for any set of open paths $B_{k}$ with
common end points we must have the positive determinant%
\begin{equation}
\det_{1\leq k,l\leq M}W(C_{kl})\geq0\,. \label{det-W-positive}%
\end{equation}
Note that in the case $M=2$ we reproduce the old inequality
(\ref{W-C-ineq-general}).

\subsection{Examples}

\label{general-example-subsection}

If we take several open paths shown in Fig.~\ref{fig-3} then we obtain the
following inequality for the rectangular $T\times R$ Wilson loops $W(T,R)$%
\begin{equation}
\det_{1\leq k,l\leq M}W\left(  \frac{T_{k}+T_{l}}{2},R\right)  \geq0\,.
\label{rect-ineq-multipoint}%
\end{equation}
This inequality should hold for any finite set of points $T_{k}$.

\begin{figure}
[ptb]
\begin{center}
\includegraphics[
height=4.2955in,
width=4.6683in
]%
{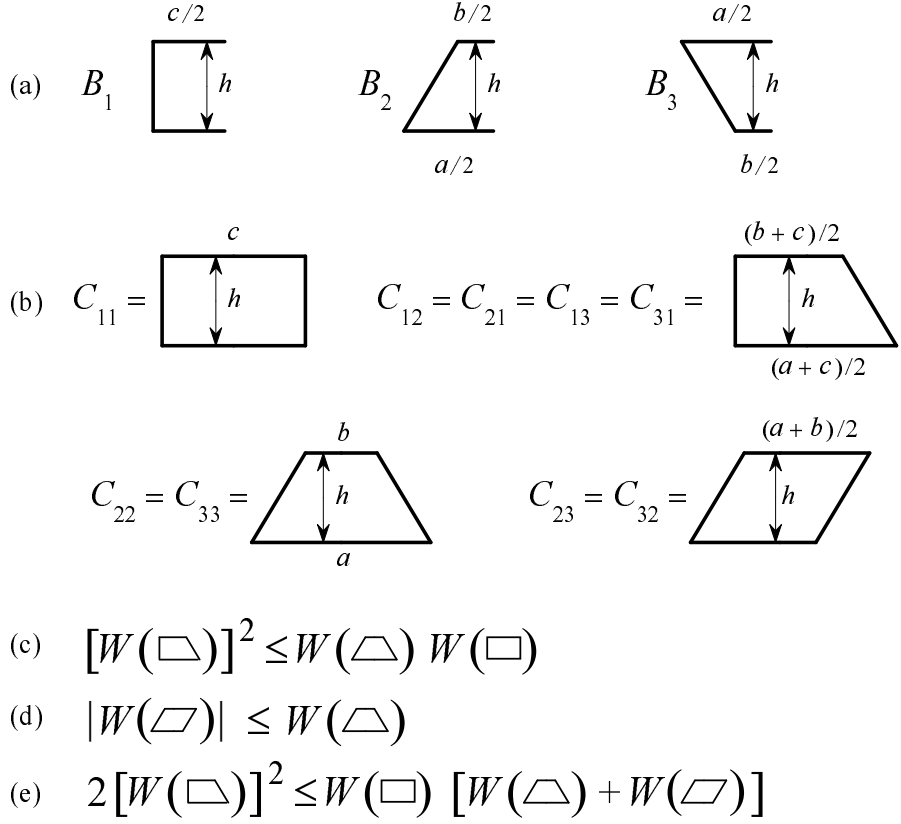}%
\caption{(a) Open paths $B_{k}$ used for the derivation of inequalities. (b)
Corresponding contours $C_{kl}$. (c), (d), (e) Graphic representation of
inequalities (\ref{rt-x-ineq}), (\ref{t-p-ineq}), (\ref{rtp-x-ineq}).}%
\label{fig-6}%
\end{center}
\end{figure}

Concerning applications of inequality (\ref{det-W-positive}) to nonrectangular
contours, let us consider the simple example corresponding to open paths
$B_{k}$ shown in Fig.~\ref{fig-6} (a).%

Let us introduce a compact notation for the contours $C_{kl}$ shown in Fig.
\ref{fig-6} (b):%
\begin{align}
W(C_{11})  &  \equiv r\quad\mathrm{(rectangle)}\,,\\
W(C_{22})  &  =W(C_{33})\equiv t\quad\mathrm{(trapezium)}\,,\\
W(C_{23})  &  =W(C_{32})\equiv p\quad\mathrm{(parallelogram)}\,,\\
W(C_{12})  &  =W(C_{13})=W(C_{21})=W(C_{31})\equiv x\,.
\end{align}
Then the positivity of the matrix%
\begin{equation}
W(C_{kl})=\left(
\begin{array}
[c]{ccc}%
r & x & x\\
x & t & p\\
x & p & t
\end{array}
\right)  \label{W-3-matrix}%
\end{equation}
leads to the positivity of the following minor determinants%
\begin{align}
\det(r)  &  =r\geq0\,,\\
\det(r)  &  =t\geq0\,,\\
\det\left(
\begin{array}
[c]{cc}%
r & x\\
x & t
\end{array}
\right)   &  =rt-x^{2}\geq0\,,\\
\det\left(
\begin{array}
[c]{cc}%
t & p\\
p & t
\end{array}
\right)   &  =t^{2}-p^{2}\geq0\,,
\end{align}
which results in%
\begin{align}
x^{2}  &  \leq rt\,,\label{rt-x-ineq}\\
|p|  &  \leq t\,. \label{t-p-ineq}%
\end{align}
The positivity of the main determinant%
\begin{equation}
\det\left(
\begin{array}
[c]{ccc}%
r & x & x\\
x & t & p\\
x & p & t
\end{array}
\right)  =r(t^{2}-p^{2})-2x^{2}(t-p)>0
\end{equation}
results in%
\begin{equation}
r(t-p)(t+p)\geq2x^{2}(t-p)\,.
\end{equation}
If $t\neq p$ then according to inequality (\ref{t-p-ineq}) we can cancel $t-p$%
\begin{equation}
2x^{2}\leq r(t+p)\,. \label{rtp-x-ineq}%
\end{equation}
But actually this inequality holds also at $t=p$ [in this case one has to use
inequality (\ref{rt-x-ineq})].

We can restrict ourselves to the set (\ref{t-p-ineq}), (\ref{rtp-x-ineq}).
Inequality (\ref{rt-x-ineq}) is their consequence. Graphically inequalities
(\ref{rt-x-ineq}), (\ref{t-p-ineq}), (\ref{rtp-x-ineq}) are shown in Fig.
\ref{fig-6} (c,d,e). Certainly the above derivation of these inequalities from
the positivity of the matrix (\ref{W-3-matrix}) is not the shortest way. We
simply wanted to illustrate how the general method based on the positivity of
matrix $W(C_{kl})$ works.

\subsection{Area law}

\label{general-area-law-subsection}

Note that inequalities (\ref{det-W-positive}) have many basic properties which
were established earlier for the simple inequality (\ref{W-C-ineq-general}).
For example, these inequalities are again compatible with the area law.
Indeed, keeping in mind that length $L(C_{kl})$ and area $S(C_{kl})$ of
$C_{kl}$ are additive in $B_{k}$ and $B_{l}$ for flat contours\footnote{In the
case of non-flat contours, the additivity of the surface area can break down
as shown in Sec.~\ref{Further-generalizations-subsection}}%
\begin{align}
L(C_{kl}) &  =L(B_{k})+L(B_{l})\,,\label{L-C-via-B}\\
S(C_{kl}) &  =S(B_{k})+S(B_{l})\,,\label{S-additivity}%
\end{align}
we find in the case of the area law (\ref{area-law-general}) for large
contours:%
\begin{align}
W(C_{kl}) &  =A(C_{kl})\,\exp\left[  -a_{0}L(C_{kl})-KS(C_{kl})\right]
\nonumber\\
&  =A(C_{kl})\,\exp\left[  -a_{0}L(B_{k})-KS(B_{k})\right]  \exp\left[
-a_{0}L(B_{l})-KS(B_{l})\right]  \,.
\end{align}
Therefore%
\begin{equation}
\det_{1\leq k,l\leq M}W(C_{kl})=\left\{  \prod\limits_{k=1}^{M}\exp\left[
-a_{0}L(B_{k})-KS(B_{k})\right]  \right\}  ^{2}\det_{1\leq k,l\leq M}%
A(C_{kl})\,.
\end{equation}
Thus inequality (\ref{det-W-positive}) takes the form%
\begin{equation}
\det_{1\leq k,l\leq M}A(C_{kl})\geq0\,.\label{det-A-general}%
\end{equation}
We see that the exponentially growing area and perimeter terms are factored
out from the inequality.

\subsection{Renormalization}

\label{renormalization-general-subsection}

Now we want to study the renormalization of the inequality
(\ref{det-W-positive}) which was derived for the nonrenormalized Wilson loops%
\begin{equation}
\det_{1\leq k,l\leq M}W^{\mathrm{nonren}}(C_{kl})\geq0\,.
\label{det-nonreg-ineq}%
\end{equation}

The results of our analysis will be quite similar to results obtained in Sec.
\ref{survival-subsection} for the case $M=2$:

1) If all open paths $B_{k}$ have the same external cusp angles%
\begin{align}
\gamma_{1}  &  =\gamma_{2}=\ldots=\gamma_{M}\equiv\gamma
\,,\label{gamma-external}\\
\gamma_{1}^{\prime}  &  =\gamma_{2}^{\prime}=\ldots=\gamma_{M}^{\prime}%
\equiv\gamma^{\prime} \label{gamma-external-prime}%
\end{align}
then all $n!$ terms of the determinant (\ref{det-nonreg-ineq}) have the same
renormalization constant so that the renormalized version of the inequality is
simply%
\begin{equation}
\det_{1\leq k,l\leq M}W^{\mathrm{ren}}(C_{kl})\geq0\,. \label{det-W-gen-ren}%
\end{equation}

2) In the case when conditions (\ref{gamma-external}),
(\ref{gamma-external-prime}) are violated, the renormalization of the
inequality (\ref{det-nonreg-ineq}) does not lead to new inequalities. In this
case one arrives either at trivial inequalities or at inequalities which can
be directly derived from the inequalities obtained in the case
(\ref{gamma-external}), (\ref{gamma-external-prime}).

These properties are proved in Appendix \ref{Appendix-section}.

\subsection{Further generalizations}

\label{Further-generalizations-subsection}

\begin{figure}
[ptb]
\begin{center}
\includegraphics[
height=4.0171in,
width=4.152in
]%
{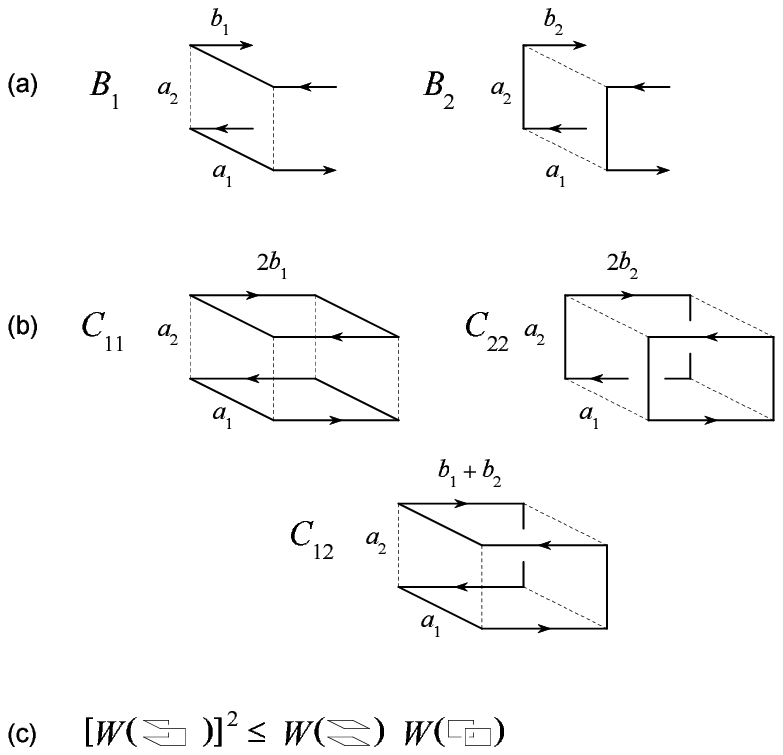}%
\caption{(a) Open paths $B_{k}$ with four end points. (b) Corresponding
contours $C_{kl}$. (c) Graphic representation of the inequality.}%
\label{fig-7}%
\end{center}
\end{figure}

So far we used connected open paths $B_{k}$ with two end points for the
derivation of inequalities. In Fig.~\ref{fig-7} (a) we show another example
with four end points. This choice leads to contours $C_{kl}$ shown in Fig.
\ref{fig-7} (b). The general inequality (\ref{W-C-ineq-general}) for these
contours $C_{kl}$ is shown graphically in Fig.~\ref{fig-7} (c).%

The check of the compatibility of this inequality with the area law is tricky
because now we cannot use the additivity of the surface area
(\ref{S-additivity}). In fact, we must work with the minimal surface. In the
case of contours $C_{kl}$ shown in Fig.~\ref{fig-7} (b) the minimal surface
area is given by
\begin{align}
S_{\min}(C_{11})  &  =\min\left[  4a_{1}b_{1},2\left(  a_{1}+2b_{1}\right)
a_{2}\right]  \,,\nonumber\\
S_{\min}(C_{22})  &  =\min\left[  4a_{2}b_{2},2\left(  a_{2}+2b_{2}\right)
a_{1}\right]  \,,\nonumber\\
S_{\min}(C_{12})  &  =a_{1}a_{2}+2(b_{1}+b_{2})\min(a_{1},a_{2})\,.
\end{align}
Using the general inequality%
\begin{equation}
\min(u,v)+\min(x,y)\leq\min(u+y,x+v)\,,
\end{equation}
we find%

\begin{align}
&  S_{\min}(C_{11})+S_{\min}(C_{22})\nonumber\\
&  =\min\left[  4a_{1}b_{1},2\left(  a_{1}+2b_{1}\right)  a_{2}\right]
+\min\left[  4a_{2}b_{2},2\left(  a_{2}+2b_{2}\right)  a_{1}\right]
\nonumber\\
&  \leq\min\left\{  \left[  4a_{1}b_{1}+2\left(  a_{2}+2b_{2}\right)
a_{1}\right]  ,\left[  4a_{2}b_{2}+2\left(  a_{1}+2b_{1}\right)  a_{2}\right]
\right\} \nonumber\\
&  =2a_{1}a_{2}+4\min\left[  \left(  a_{1}b_{1}+a_{1}b_{2}\right)  ,\left(
a_{2}b_{1}+a_{2}b_{2}\right)  \right] \nonumber\\
&  =2a_{1}a_{2}+4\left(  b_{1}+b_{2}\right)  \min\left(  a_{1},a_{2}\right)
=2S_{\min}(C_{12})\,.
\end{align}
Thus%
\begin{equation}
S_{\min}(C_{11})+S_{\min}(C_{22})\leq2S_{\min}(C_{12})\,,
\end{equation}
which proves the compatibility of inequality (\ref{W-C-ineq-general}) with the
area law.

\subsection{Positivity and spectral representation for rectangular Wilson loops}

\label{Positivity-spectral-subsection}

As was already mentioned, inequalities for rectangular Wilson loops [the
original Bachas inequality (\ref{Bachas-ineq}) and its generalization
(\ref{rect-ineq-multipoint})] can be considered as particular cases of the
general inequality (\ref{det-W-positive}). However, in the rectangular case we
have much better understanding of the positivity constraints. It comes from
the spectral representation for rectangular Wilson loops. Using the gauge
$A_{4}=0$, we can write%
\begin{align}
W(T,R) &  =\frac{1}{N}\sum\limits_{ab}\langle0|\left[  U_{ab}(0,R)\right]
^{+}e^{-HT}U_{ab}(0,R)|0\rangle\nonumber\\
&  =\frac{1}{N}\sum\limits_{ab}\sum\limits_{n}e^{-E_{n}(R)T}\left|  \langle
n|U_{ab}(0,R)|0\rangle\right|  ^{2}\,.
\end{align}
This can be rewritten in the form%
\begin{equation}
W(T,R)=\int_{E_{0}(R)}^{\infty}dEe^{-ET}\rho(E,R)\label{W-spectral}%
\end{equation}
where%
\begin{equation}
\rho(E,R)=\frac{1}{N}\sum\limits_{ab}\sum\limits_{n}\delta(E-E_{n}%
(R))e^{-E_{n}(R)T}\left|  \langle n|U_{ab}(0,R)|0\rangle\right|  ^{2}\,.
\end{equation}
Obviously%
\begin{equation}
\rho(E,R)\geq0\,.
\end{equation}
Starting from the spectral representation (\ref{W-spectral}), one can easily
reproduce inequality (\ref{rect-ineq-multipoint}). Indeed,%
\begin{equation}
\sum\limits_{kl}x_{k}x_{l}W\left(  \frac{T_{k}+T_{l}}{2},R\right)
=\int_{E_{0}(R)}^{\infty}dE\left(  \sum\limits_{k}x_{k}e^{-ET_{k}/2}\right)
^{2}\rho(E,R)\geq0\,.
\end{equation}
This shows that the quadratic form $W\left(  \left(  T_{k}+T_{l}\right)
/2,R\right)  $ is positive definite, which results in inequality
(\ref{rect-ineq-multipoint}).

\section{Inequalities and cusp divergences in the effective string model for
Wilson loops}

\label{String-section}

\setcounter{equation}{0} 

\subsection{Effective string model}

In Refs. \cite{Luescher-80}, \cite{Luescher-81} it was suggested to consider
Nambu-Goto string with the boundary fixed on a closed contour $C$ as an
effective model for Wilson loops $W(C)$ with large smooth contours $C$. In
this model the Wilson loop $W(C)$ is approximated%
\begin{equation}
W_{\mathrm{LSW}}(C)\sim\mathcal{Z}(C)\, \label{W-z-model}%
\end{equation}
by the string ``partition function'' $\mathcal{Z}(C)$ in $D$-dimensional
space-time computed for open strings fixed with Dirichlet boundary conditions
(up to a reparametrization) on contour $C$. For flat contours $C$ this
partition function (computed in the quadratic approximation in deviations from
the minimal surface) reduces to the determinant of Laplace operator
$\Delta_{C}$ with Dirichlet boundary conditions on contour $C$%
\begin{equation}
\mathcal{Z}(C)\sim e^{-KS(C)}\left[  \mathrm{Det}\,(-\Delta_{C})\right]
^{-(D-2)/2}\,. \label{z-model-det}%
\end{equation}

Applying this model to (smoothed) rectangular contours, one can extract
\cite{Luescher-80}, \cite{Luescher-81} the $1/R$ correction (\ref{V-large-R})
to the potential (\ref{V-large-R}) at large distances:%
\begin{equation}
V(R)\overset{R\rightarrow\infty}{=}KR+2a_{0}-\frac{\pi(D-2)}{24}\frac{1}%
{R}\,.\label{V-Luscher}%
\end{equation}
Without touching the interesting question about the theoretical status of the
LSW model, we would like to concentrate on the consistency of the model with
the inequalities derived in this paper for Wilson loops.

\subsection{Cusp singularities}

\label{Cusp-string-subsection}

The small-$t$ expansion (\ref{Gamma-cusp-string}) leads to the factorized
structure of the ultraviolet divergences for the polygonal contours%
\begin{align}
\left[  \mathrm{Det}\,(-\Delta_{C})\right]  _{\mathrm{ren}} &  =\lim
_{\Lambda\rightarrow\infty}\left[  \exp\left[  \mathrm{const}\,S(C)\Lambda
^{2}+\,\mathrm{const}\,L(C)\Lambda\right]  \prod_{i}\Lambda^{-2\Gamma
_{\mathrm{cusp}}^{\mathrm{str}}(\theta_{i})}\right]  \nonumber\\
&  \times\left[  \mathrm{Det}\,(-\Delta_{C})\right]  _{\mathrm{nonren}%
,\Lambda}\,.\label{Z-string-total}%
\end{align}

The renormalization constant appearing in Eq.~(\ref{Z-string-total}) should be
taken to the power $-(D-2)/2$ according to Eq.~(\ref{z-model-det}). Comparing
the string (\ref{Z-string-total}) and gauge (\ref{Z-cusp-large-Lambda}) cusp
factors%
\begin{align}
\mathrm{Gauge\,theories} &  :\quad Z_{\mathrm{cusp}}(\Lambda,\gamma
)\overset{\Lambda\rightarrow\infty}{\sim}\left[  g(\Lambda)\right]
^{-\Gamma(\gamma)/\beta_{1}}\sim\left[  \ln\Lambda\right]  ^{\Gamma
(\gamma)/(2\beta_{1})}\,,\label{Z-cusp-gauge}\\
\mathrm{String\,model} &  :\quad Z_{\mathrm{cusp}}(\Lambda,\gamma
)\overset{\Lambda\rightarrow\infty}{\sim}\Lambda^{(D-2)\Gamma_{\mathrm{cusp}%
}^{\mathrm{str}}(\theta_{i})}\,,\label{Z-cusp-string}%
\end{align}
we see that the dependence on $\Lambda\;$is different. Therefore the direct
comparison of Wilson loops $W(C)$ and string functionals $\mathcal{Z}(C)$ for
contours with cusps makes no sense. In fact, the physical assumptions standing
behind the effective string model also suggest that this model may be relevant
only for large smooth contours. On the other hand, one can construct ratios of
Wilson loops where cusp singularities cancel. For these ratios the comparison
between gauge theories and the string model is still possible. In order to
avoid uncertainties related to the area and perimeter dependent divergences,
the ratios of string functionals $\mathcal{Z}(C)$ should be organized so that
the exponential perimeter and area dependences cancel.

Thus we want to concentrate on ratios of Wilson lines $W(C)$ (and ratios of
corresponding string functionals $\mathcal{Z}(C)$) for which

1) the perimeter and area exponential factors cancel,

2) the full set of cusp angels associated with the numerator of the ratio
coincides with the full set of the denominator angles.

But these are exactly the properties of the ratios of Wilson loops which
appear in the inequalities discussed in this article. Now one can ask the
question whether these inequalities hold in the string model. A serious
analysis of this problem is beyond the scope of this paper. But several
arguments in favor of this possibility can be suggested now.

First, the string cusp anomalous dimension (\ref{Gamma-cusp-string}) obeys
convexity inequality (\ref{Gamma-1-convex-string}), which is a trivial
consequence of the negative second derivative $d^{2}\Gamma_{\mathrm{cusp}%
}^{\mathrm{str}}(\theta)/d\theta^{2}<0$. This is an analog of the convexity
property of the cusp anomalous dimension in gauge theories
(\ref{Gamma-1-convex-string}). If polygon inequalities hold in the string
model then the property (\ref{Gamma-1-convex-string}) would keep the
consistency of inequalities with respect to the renormalization of cusp
singularities. Although the $\Lambda$ dependence of cusp renormalization
factors (\ref{Z-cusp-gauge}), (\ref{Z-cusp-string}) is different (logarithmic
in gauge theories and power in the string model) the qualitative features are
the same: both $\Lambda$ and $\ln\Lambda$ grow to infinity when $\Lambda
\rightarrow\infty$. Therefore the stability of inequalities under the
renormalization is completely controlled by the convexity properties which are
the same for the gauge and string anomalous dimensions. Strictly speaking,
there is one small difference. The gauge anomalous dimension is invariant
under $\gamma\rightarrow-\gamma$ whereas the corresponding transformation
$\theta\rightarrow2\pi-\theta$ changes the string anomalous dimension (the
inside polygon angle $\theta$ is implied in (\ref{Gamma-cusp-string})). This
is the reason why we prefer to write $\Gamma_{\mathrm{cusp}}^{\mathrm{str}}$
as a function of the inside angle $\theta$ ($0<\theta<2\pi$) and not the
deviation angle $\gamma=\pi-\theta$ as in the case of the gauge theory. But
one can show that the absence of the $\gamma\rightarrow-\gamma$ symmetry does
not break the arguments of Secs. \ref{triangle-limit-subsection},
\ref{survival-subsection} and the same logic can be applied to the string model.

\subsection{Rectangular contour}

\label{Rectangular-string-subsection}

Another evidence for the validity of inequalities comes from the analysis of
the simple rectangular case. The results for rectangular loops discussed here
are well known \cite{Dietz-83}. We use them in order to illustrate how
inequalities (\ref{W-z-model}), (\ref{z-model-det}) work in the effective
string model.

In the case of the rectangular $T\times R$ contour, the cusp renormalization
factor generated by four angles $\theta_{i}=\pi/2$ and appearing in Eq.
(\ref{Z-string-total}) equals according to Eq.~(\ref{Gamma-cusp-string})
\begin{equation}
\prod_{i=1}^{4}\Lambda^{-2\Gamma_{\mathrm{cusp}}^{\mathrm{str}}(\theta_{i}%
)}=\Lambda^{-8\Gamma_{\mathrm{cusp}}^{\mathrm{str}}(\pi/2)}=\Lambda^{1/2}\,.
\end{equation}
Therefore%
\begin{align}
\mathcal{Z}_{\mathrm{ren}}(T,R) &  =\lim_{\Lambda\rightarrow\infty}\left\{
\Lambda^{1/2}\exp\left[  \mathrm{const}\,S(C)\Lambda^{2}+\,\mathrm{const}%
\,L(C)\Lambda\right]  \right\}  ^{-(D-2)/2}\nonumber\\
&  \times\mathcal{Z}_{\mathrm{nonren}}(\Lambda,T,R)\,.
\end{align}
Combining this with the dimensional counting, we see that%
\begin{equation}
\mathcal{Z}_{\mathrm{ren}}(T,R)=e^{-KR-2a_{0}\left(  T+R\right)  }\left[
\left(  RT\right)  ^{1/4}f\left(  T/R\right)  \right]  ^{(D-2)/2}%
\end{equation}
where function $f$ depends only on the ratio $T/R$. Function $f\left(
T/R\right)  $ was computed in Ref. \cite{Dietz-83} and the result
is\footnote{In the case of general polygonal contours $C$ the deteminant of
Laplace operator $\Delta_{C}$ was studied in Ref. \cite{AS-93}.} (we omit
constant factors depending on the renormalization scheme)%
\begin{equation}
\mathcal{Z}_{\mathrm{ren}}(T,R)=e^{-KRT-2a_{0}\left(  T+R\right)  }\left[
R^{-1/2}\eta\left(  iT/R\right)  \right]  ^{-(D-2)/2}\,.\label{z-ren-eta}%
\end{equation}
Here $\eta(\tau)$ is Dedekind eta function (\ref{eta-infinite-product}) whose
properties are briefly described in Appendix \ref{Dedekind-appendix}.
Expression (\ref{z-ren-eta}) also agrees with the determinant of the lattice
regularized Laplacian computed in Ref. \cite{DD-88}.

Using the property (\ref{eta-inv}) of the $\eta$ function, one can see that
$\mathcal{Z}_{\mathrm{ren}}(T,R)$ is symmetric%
\begin{equation}
\mathcal{Z}_{\mathrm{ren}}(T,R)=\mathcal{Z}_{\mathrm{ren}}(R,T)\,
\end{equation}
as it should be.

Taking the limit of large $T\rightarrow\infty$ in the rectangular contour of
Eq.~(\ref{z-ren-eta}), one can easily reproduce L\"{u}scher term
(\ref{V-Luscher}). Indeed, using Eq.~(\ref{eta-tau-large}), we find from Eq.
(\ref{z-ren-eta})
\begin{equation}
\mathcal{Z}_{\mathrm{ren}}(T,R)\overset{T\rightarrow\infty}{=}e^{-KRT-2a_{0}%
\left(  T+R\right)  }\left[  R^{-1/2}e^{-(\pi/12)T/R}\right]  ^{-(D-2)/2}\,.
\end{equation}
Combining this with Eqs. (\ref{V-via-W}) and (\ref{W-z-model}), one arrives at
Eq.~(\ref{V-Luscher}).

\subsection{Spectral decomposition and inequalities for rectangular contours}

\label{Spectral-string-section}

Inserting the power series for $\left[  \eta(i\tau)\right]  ^{-\alpha}$
(\ref{eta-alpha-pos}) into Eq.~(\ref{z-ren-eta}), one finds%
\begin{align}
\mathcal{Z}_{\mathrm{ren}}(T,R) &  =e^{-KRT-2a_{0}\left(  T+R\right)  }\left[
R^{-1/2}\eta\left(  iT/R\right)  \right]  ^{-(D-2)/2}\nonumber\\
&  =e^{-KRT-2a_{0}\left(  T+R\right)  }\left[  R\,e^{(\pi/6)T/R}\right]
^{(D-2)/4}\sum\limits_{n=0}^{\infty}A_{n}\left(  \frac{D-2}{2}\right)
e^{-2\pi nT/R}\,.\label{z-ren-res}%
\end{align}
where according to (\ref{A-n-positive})%
\begin{equation}
A_{n}\left(  \frac{D-2}{2}\right)  >0\,.
\end{equation}
Eq.~(\ref{z-ren-res}) can be rewritten in the form%
\begin{equation}
\mathcal{Z}_{\mathrm{ren}}(T,R)=\sum\limits_{n=0}C_{n}(R)e^{-E_{n}%
(R)T}\label{z-spectral}%
\end{equation}
where%
\begin{align}
E_{n}(R) &  =KR+2a_{0}+\left(  -\frac{D-2}{24}+n\right)  \frac{\pi}%
{R}\,,\label{E-n-string}\\
C_{2n}(R) &  =e^{-2a_{0}R}R^{(D-2)/4}A_{n}\left(  \frac{D-2}{2}\right)
>0\,,\\
C_{2n+1}(R) &  =0\,.
\end{align}
The full spectrum is twice larger than can be seen in expansion
(\ref{z-ren-res}) because the odd coefficients $C_{2n+1}(R)$ vanish. The lost
part of the spectrum $E_{n}(R)$ (\ref{E-n-string}) can be found if one
considers the partition function of the string model%
\begin{equation}
\mathcal{Z}_{\beta}(R)=\sum\limits_{n=0}w_{n}e^{-\beta E_{n}(R)}%
\label{Z-beta-general}%
\end{equation}
where $\beta$ is inverse temperature and $w_{n}$ are integer degeneracy
weights of levels $E_{n}(R)$. Since partition function $\mathcal{Z}_{\beta
}(R)$ corresponds to periodic boundary conditions in the functional integral,
one has to modify the boundary conditions in Eq.~(\ref{z-model-det})%
\begin{equation}
\mathcal{Z}_{\beta}(R)\sim\left[  \mathrm{Det}\,(-\Delta_{\mathrm{mixed}%
})\right]  ^{-(D-2)/2}\,,
\end{equation}%
\begin{equation}
\Delta_{\mathrm{mixed}}:\quad\left\{
\begin{array}
[c]{cc}%
T & \mathrm{periodic}\\
R & \mathrm{Dirichlet}%
\end{array}
\right.  \,.
\end{equation}
The result of Ref. \cite{Dietz-83} for these boundary conditions is%
\begin{equation}
\mathcal{Z}_{\beta}(R)=\left[  \eta\left(  i\frac{\beta}{2R}\right)  \right]
^{-(D-2)}\,.\label{Z-partition-function}%
\end{equation}
Using expansion (\ref{eta-alpha-pos}) for the $\eta$ function, one finds%
\begin{equation}
\mathcal{Z}_{\beta}(R)=e^{-KRT-2a_{0}\left(  T+R\right)  }e^{\beta
(D-2)\pi/(24R)}\sum\limits_{n=0}^{\infty}A_{n}(D-2)e^{-\pi n\beta/R}\,.
\end{equation}
Comparing this with (\ref{Z-beta-general}), one obtains spectrum
(\ref{E-n-string}) and the degeneracies%
\begin{equation}
w_{n}=A_{n}(D-2)\,.
\end{equation}
Note that numbers $A_{n}(D-2)$ are integer (see Eq.~(\ref{A-n-m-via-P-n}) in
Appendix \ref{Dedekind-appendix}).

Series (\ref{z-spectral}) is a spectral decomposition of type
(\ref{W-spectral}) with the spectral density made of discrete delta-functions.
As was shown in section \ref{Positivity-spectral-subsection}, representation
(\ref{W-spectral}) guarantees that inequality (\ref{rect-ineq-multipoint})
holds. Thus the string model for Wilson loops (\ref{W-z-model}),
(\ref{z-model-det}) satisfies inequalities (\ref{rect-ineq-multipoint}) for
rectangular contours $C$.

In terms of two-dimensional Laplacians $\Delta(T,R)$ with Dirichlet boundary
conditions on $T\times R$ rectangular contours these inequalities can be
rewritten in the form%
\begin{equation}
\det_{1\leq k,l\leq M}\left\{  \left[  \mathrm{Det}\,\Delta\left(
\frac{T_{k}+T_{l}}{2},R\right)  \right]  ^{-\nu}\right\}  \geq0\,
\label{Det-Delta-rect-ineq}%
\end{equation}
where $\nu>0$.

\section{Conclusions}

\setcounter{equation}{0} 

We have considered two ways of the generalization of the original Bachas
inequality (\ref{Bachas-ineq}):

1) extension to nonrectangular contours (\ref{W-C-ineq-general}),

2) determinant inequalities (\ref{det-W-positive}) containing several
composite contours.

The transition from the rectangular contours to arbitrary polygons does not
allow us to use the lattice regularization (at least its standard cubic
version). Therefore one should take a special care about the renormalization
of inequalities. We have found that naive polygon inequalities survive in the
renormalized theory only if the balance of cusp angles (\ref{cusp-balance})
holds. In the absence of this balance the inequalities become trivial in the
continuum limit but still correct. This correctness follows from the convexity
property of cusp anomalous dimensions (\ref{Gamma-convex}).

In this paper we used polygonal contours in order to concentrate on the
properties of cusp singularities. The general inequality (\ref{det-W-gen-ren})
also holds for curved contours with cusps or without them.

Since the inequalities are too general and hold in any gauge theory, the area
law cannot be derived from these inequalities without using an additional
dynamic input. Nevertheless it is interesting that the inequalities are
compatible with the area law and the area and perimeter exponential terms can
be factored out from the inequalities.

Our analysis of the inequalities for the determinants of two-dimensional
Laplacians is incomplete. Only rectangular inequalities
(\ref{Det-Delta-rect-ineq}) were checked. Although the discussion was
presented in the context of the LSW model but actually one deals with a rather
interesting pure mathematical problem.


\textbf{Acknowledgements.} I am grateful to E. Antonov, I.~Cherednikov,
D.I.~Diakonov, M. Eides, N.~Kivel, V. Kudryavtsev, A. Losev, V.Yu.~Petrov,
M.V.~Polyakov and N.G.~Stefanis for useful discussions. This work was
supported by DFG and BMBF.

\appendix                                                  

\renewcommand{\theequation}{\Alph{section}.\arabic{equation}} 

\section{Combinatorics of the renormalization}

\setcounter{equation}{0} 

\label{Appendix-section}

In this appendix we prove the renormalization properties of inequality
(\ref{det-nonreg-ineq}) formulated in Sec.
\ref{renormalization-general-subsection}.

Using general renormalization equations (\ref{W-renormalization}),
(\ref{Z-per-cusp}), we reduce inequality (\ref{det-nonreg-ineq}) to%
\begin{equation}
\det_{1\leq k,l\leq M}\left[  Z_{\mathrm{per.}}^{-1}(\Lambda,L(C_{kl}%
))\prod_{C_{kl}\,\mathrm{cusps}}Z_{\mathrm{cusp}}^{-1}(\Lambda,\gamma
)W^{\mathrm{ren}}(C_{kl})\right]  \geq0\,.
\end{equation}
Combining Eqs. (\ref{Z-per-mult}) and (\ref{L-C-via-B}), we find%
\begin{equation}
Z_{\mathrm{per.}}(\Lambda,L(C_{kl}))=Z_{\mathrm{per.}}(\Lambda,L(B_{k}%
))Z_{\mathrm{per.}}(\Lambda,L(B_{l}))\,.
\end{equation}
Hence%
\begin{align}
&  \det_{1\leq k,l\leq M}\left\{  Z_{\mathrm{per.}}^{-1}(\Lambda
,L(C_{kl}))\left[  \prod_{C_{kl}\,\mathrm{cusps}}Z_{\mathrm{cusp}}%
^{-1}(\Lambda,\gamma)\right]  W^{\mathrm{ren}}(C_{kl})\right\} \nonumber\\
&  =\left[  \prod_{k=1}^{M}Z_{\mathrm{per.}}(\Lambda,L(B_{k}))\right]
^{-2}\det_{1\leq k,l\leq M}\left\{  \left[  \prod_{C_{kl}\,\mathrm{cusps}%
}Z_{\mathrm{cusp}}^{-1}(\Lambda,\gamma)\right]  W^{\mathrm{ren}}%
(C_{kl})\right\}
\end{align}
and inequality (\ref{det-nonreg-ineq}) reduces to%
\begin{equation}
\det_{1\leq k,l\leq M}\left\{  \left[  \prod_{C_{kl}\,\mathrm{cusps}%
}Z_{\mathrm{cusp}}^{-1}(\Lambda,\gamma)\right]  W^{\mathrm{ren}}%
(C_{kl})\right\}  \geq0\,. \label{det-ren-ineq-1}%
\end{equation}

If path $B_{k}$ has external cusp angles $\gamma_{k}$, $\gamma_{k}^{\prime}$
(see Fig.~\ref{fig-5}) then%

\begin{align}
\prod_{C_{kl}\,\mathrm{cusps\,}}Z_{\mathrm{cusp}}^{-1}(\Lambda,\gamma)  &
=\left[  \prod_{B_{k}\,\mathrm{cusps}}Z_{\mathrm{cusp}}^{-1}(\Lambda
,\gamma)\right]  \left[  \prod_{B_{l}\,\mathrm{cusps}}Z_{\mathrm{cusp}}%
^{-1}(\Lambda,\gamma)\right] \nonumber\\
&  \times Z_{\mathrm{cusp}}^{-1}(\Lambda,\gamma_{k}+\gamma_{l}%
)Z_{\mathrm{cusp}}^{-1}(\Lambda,\gamma_{k}^{\prime}+\gamma_{l}^{\prime})\,.
\label{Gamma-cusp-decomposition-1}%
\end{align}
Therefore%
\begin{align}
&  \det_{1\leq k,l\leq M}\left\{  \left[  \prod_{C_{kl}\,\mathrm{cusps}%
}Z_{\mathrm{cusp}}^{-1}(\Lambda,\gamma)\right]  W^{\mathrm{ren}}%
(C_{kl})\right\}  =\left[  \prod_{B_{k}\,\mathrm{cusps}}Z_{\mathrm{cusp}}%
^{-1}(\Lambda,\gamma)\right]  ^{2}\nonumber\\
&  \times\det_{1\leq k,l\leq M}\left\{  Z_{\mathrm{cusp}}^{-1}(\Lambda
,\gamma_{k}+\gamma_{l})Z_{\mathrm{cusp}}^{-1}(\Lambda,\gamma_{k}^{\prime
}+\gamma_{l}^{\prime})W^{\mathrm{ren}}(C_{kl})\right\}
\end{align}
and inequality (\ref{det-ren-ineq-1}) simplifies to%
\begin{equation}
\det_{1\leq k,l\leq M}\left\{  Z_{\mathrm{cusp}}^{-1}(\Lambda,\gamma
_{k}+\gamma_{l})Z_{\mathrm{cusp}}^{-1}(\Lambda,\gamma_{k}^{\prime}+\gamma
_{l}^{\prime})W^{\mathrm{ren}}(C_{kl})\right\}  \geq0\,.
\label{det-ren-ineq-2}%
\end{equation}

Now we have to consider different cases separately. In the simplest case when
all open paths $B_{k}$ have equal external cusp angles (\ref{gamma-external}),
(\ref{gamma-external-prime}) we arrive at%
\begin{align}
&  \det_{1\leq k,l\leq M}\left\{  Z_{\mathrm{cusp}}^{-1}(\Lambda,\gamma
_{k}+\gamma_{l})Z_{\mathrm{cusp}}^{-1}(\Lambda,\gamma_{k}^{\prime}+\gamma
_{l}^{\prime})W^{\mathrm{ren}}(C_{kl})\right\} \nonumber\\
&  =\left[  Z_{\mathrm{cusp}}^{-1}(\Lambda,2\gamma)Z_{\mathrm{cusp}}%
^{-1}(\Lambda,2\gamma^{\prime})\right]  ^{M}\det_{1\leq k,l\leq M}\left\{
W^{\mathrm{ren}}(C_{kl})\right\}
\end{align}
and inequality (\ref{det-ren-ineq-2}) becomes%
\begin{equation}
\det_{1\leq k,l\leq M}W^{\mathrm{ren}}(C_{kl})\geq0\,. \label{det-ren-res}%
\end{equation}
Thus in the case (\ref{gamma-external}), (\ref{gamma-external-prime}) the
renormalized inequality (\ref{det-ren-res}) has the same form as the
nonrenormalized inequality (\ref{det-nonreg-ineq}). This completes the
derivation of inequality (\ref{det-W-gen-ren}) announced in Sec.
\ref{renormalization-general-subsection}.

As was already announced in Sec.~\ref{renormalization-general-subsection},
inequalities (\ref{det-ren-res}) derived under assumptions
(\ref{gamma-external}), (\ref{gamma-external-prime}) provide all information:
in the case when paths $B_{k}$ do not obey cusp balance conditions
(\ref{gamma-external}), (\ref{gamma-external-prime}) the renormalization of
inequality (\ref{det-nonreg-ineq}) will not lead to new inequalities. Let us
prove this statement.

In this case we must return to inequality (\ref{det-ren-ineq-2}). We can
represent the determinant as a sum over all permutations $P$ with signum
factor $\varepsilon(P)$:%

\begin{align}
&  \det_{1\leq k,l\leq M}\left[  Z_{\mathrm{cusp}}^{-1}(\Lambda,\gamma
_{k}+\gamma_{l})Z_{\mathrm{cusp}}^{-1}(\Lambda,\gamma_{k}^{\prime}+\gamma
_{l}^{\prime})W^{\mathrm{ren}}(C_{kl})\right] \nonumber\\
&  =\sum\limits_{P}\varepsilon(P)\prod_{k=1}^{M}\left[  Z_{\mathrm{cusp}}%
^{-1}(\Lambda,\gamma_{k}+\gamma_{P(k)})Z_{\mathrm{cusp}}^{-1}(\Lambda
,\gamma_{k}^{\prime}+\gamma_{P(k)}^{\prime})W^{\mathrm{ren}}\left(
C_{k,P(k)}\right)  \right]  \geq0\,. \label{det-sum-P}%
\end{align}
Now we multiply this inequality by%
\begin{equation}
\prod_{k=1}^{M}\left[  Z_{\mathrm{cusp}}(\Lambda,2\gamma_{k})Z_{\mathrm{cusp}%
}(\Lambda,2\gamma_{k}^{\prime})\right]  \,.
\end{equation}
Then%
\begin{equation}
\sum\limits_{P}\varepsilon(P)\prod_{k=1}^{M}\left[  \frac{Z_{\mathrm{cusp}%
}^{-1}(\Lambda,\gamma_{k}+\gamma_{P(k)})}{Z_{\mathrm{cusp}}^{-1}%
(\Lambda,2\gamma_{k})}\frac{Z_{\mathrm{cusp}}^{-1}(\Lambda,\gamma_{k}^{\prime
}+\gamma_{P(k)}^{\prime})}{Z_{\mathrm{cusp}}^{-1}(\Lambda,2\gamma_{k}^{\prime
})}W^{\mathrm{ren}}\left(  C_{k,P(k)}\right)  \right]  \geq0\,.
\label{Sum-P-ineq-1}%
\end{equation}
Next we have to identify the terms giving the dominant contribution at large
$\Lambda$. We will need the following property%
\begin{equation}
\lim_{\Lambda\rightarrow\infty}\prod_{k=1}^{M}\frac{Z_{\mathrm{cusp}}%
^{-1}(\Lambda,\gamma_{k}+\gamma_{P(k)})}{Z_{\mathrm{cusp}}^{-1}(\Lambda
,2\gamma_{k})}=\left\{
\begin{array}
[c]{ll}%
1 & \mathrm{if}\,\gamma_{k}=\gamma_{P(k)}\,\mathrm{\,for\,all\,}k\\
0 & \mathrm{otherwise}%
\end{array}
\right.  \,. \label{prod-Z-1-0}%
\end{equation}
The first case, when$\,\gamma_{k}=\gamma_{P(k)}$ for all $k$, is trivial.
Therefore we must concentrate on the case when $\gamma_{k}\neq\gamma_{P(k)}$
at least one value of $k$. According to Eq.~(\ref{Z-cusp-large-Lambda})%
\begin{equation}
\prod_{k=1}^{M}\frac{Z_{\mathrm{cusp}}^{-1}(\Lambda,\gamma_{k}+\gamma_{P(k)}%
)}{Z_{\mathrm{cusp}}^{-1}(\Lambda,2\gamma_{k})}\overset{\Lambda\rightarrow
\infty}{\rightarrow}\exp\left\{  \left[  \ln g^{2}(\Lambda)\right]
\sum\limits_{k=1}^{M}\left[  \Gamma\left(  \gamma_{k}+\gamma_{P(k)}\right)
-\Gamma\left(  2\gamma_{k}\right)  \right]  \right\}  \,. \label{prod-Z-M}%
\end{equation}
Taking into account that%
\begin{equation}
\sum\limits_{k=1}^{M}\Gamma\left(  2\gamma_{P(k)}\right)  =\sum\limits_{k=1}%
^{M}\Gamma\left(  2\gamma_{k}\right)  \,,
\end{equation}
we can write%
\begin{equation}
\sum\limits_{k=1}^{M}\left[  \Gamma\left(  \gamma_{k}+\gamma_{P(k)}\right)
-\Gamma\left(  2\gamma_{k}\right)  \right]  =\frac{1}{2}\sum\limits_{k=1}%
^{M}\left[  2\Gamma\left(  \gamma_{k}+\gamma_{P(k)}\right)  -\Gamma\left(
2\gamma_{k}\right)  -\Gamma\left(  2\gamma_{P(k)}\right)  \right]  \,.
\label{Gamma-sum-id}%
\end{equation}
According to inequality (\ref{Gamma-convex})%
\begin{equation}
\sum\limits_{k=1}^{M}\left[  2\Gamma\left(  \gamma_{k}+\gamma_{P(k)}\right)
-\Gamma\left(  2\gamma_{k}\right)  -\Gamma\left(  2\gamma_{P(k)}\right)
\right]  \geq0 \label{Gamma-ineq}%
\end{equation}
and the equality occurs only if $\gamma_{k}=\gamma_{P(k)}$ for all $k$.
Combining Eqs. (\ref{prod-Z-M}), (\ref{Gamma-sum-id}) and (\ref{Gamma-ineq}),
we prove (\ref{prod-Z-1-0}).

Now we can apply Eq.~(\ref{prod-Z-1-0}) to the analysis of the LHS of
inequality (\ref{Sum-P-ineq-1})%
\begin{align}
&  \prod_{k=1}^{M}\frac{\left[  Z_{\mathrm{cusp}}^{-1}(\Lambda,\gamma
_{k}+\gamma_{P(k)})Z_{\mathrm{cusp}}^{-1}(\Lambda,\gamma_{k}^{\prime}%
+\gamma_{P(k)}^{\prime})\right]  }{\left[  Z_{\mathrm{cusp}}^{-1}%
(\Lambda,2\gamma_{k})Z_{\mathrm{cusp}}^{-1}(\Lambda,2\gamma_{k}^{\prime
})\right]  }\nonumber\\
&  =\left\{
\begin{array}
[c]{ll}%
1 & \mathrm{if\,}\,\,\gamma_{k}=\gamma_{P(k)}\,,\,\,\gamma_{k}^{\prime}%
=\gamma_{P(k)}^{\prime}\,\mathrm{for\,all\,\,}k\\
0 & \mathrm{otherwise}%
\end{array}
\right.  \,.
\end{align}
Therefore inequality (\ref{det-sum-P}) takes the form%
\begin{equation}
\sum\limits_{P\in\Pi_{0}}(-1)^{\varepsilon(P)}\prod_{k=1}^{M}W^{\mathrm{ren}%
}\left(  C_{k,P(k)}\right)  \geq0\,. \label{Sum-P-0-ineq}%
\end{equation}
Here $\Pi_{0}$ is the subgroup of permutations obeying the condition%
\begin{equation}
\Pi_{0}=\left\{  P:\,\,\gamma_{k}=\gamma_{P(k)}\,,\,\gamma_{k}^{\prime}%
=\gamma_{P(k)}^{\prime}\,\mathrm{for\,all\,\,}k\right\}  \,.
\end{equation}

Let us say that two indices $k$ and $l$ belong to the same class if
$\gamma_{k}=\gamma_{l}$ and $\gamma_{k}^{\prime}=\gamma_{l}^{\prime}$. Then
the full set of indices can be divided into classes of equivalence $J_{\alpha
}$:
\begin{align}
\{1,2,,\ldots,M\}  &  =
{\textstyle\bigcup_{\alpha}}
J_{\alpha}\,,\\
J_{\alpha}\cap J_{\beta}  &  =\emptyset\quad(\alpha\neq\beta).
\end{align}
It is easy to see that subgroup $\Pi_{0}$ is made of permutations which leave
indices within their equivalence classes. Therefore the sum over $P$ in
inequality (\ref{Sum-P-0-ineq}) reduces to the product of determinants
restricted to different classes:
\begin{equation}
\sum\limits_{P\in\Pi_{0}}\varepsilon(P)\prod_{k=1}^{M}W^{\mathrm{ren}}\left(
C_{k,P(k)}\right)  =\prod_{\alpha}\det_{k,l\in J_{\alpha}}W^{\mathrm{ren}%
}(C_{kl})\,.
\end{equation}
Thus our inequality (\ref{Sum-P-0-ineq}) reduces to%
\begin{equation}
\prod_{\alpha}\det_{k,l\in J_{\alpha}}W^{\mathrm{ren}}(C_{kl})\geq0\,.
\label{Prod-alpha-ineq}%
\end{equation}
Now it remains to note that instead of disentangling the problems caused by
the violation of the cusp balance conditions (\ref{gamma-external}),
(\ref{gamma-external-prime}) we could concentrate from the very beginning on a
separate class of equivalent indices $k\in J_{\alpha}$ and consider the
corresponding set of open lines $B_{k}$ generating inequality%
\begin{equation}
\det_{k,l\in J_{\alpha}}W^{\mathrm{ren}}(C_{kl})\geq0\,. \label{C-alpha-ineq}%
\end{equation}
Inequality (\ref{Prod-alpha-ineq}) is simply the product of these elementary
inequalities which we have already derived in (\ref{det-ren-res}).

A small comment is needed about the case when the equivalence class
$J_{\alpha}$ contains only one index $m$. In this case inequality
(\ref{C-alpha-ineq}) becomes trivial:%
\begin{equation}
\det_{k,l\in J_{\alpha}}W^{\mathrm{ren}}(C_{kl})=W^{\mathrm{ren}}(C_{mm}%
)\geq0\,.
\end{equation}

\section{ Dedekind $\eta$ function}

\label{Dedekind-appendix}

\setcounter{equation}{0} 

Dedekind $\eta$ function \cite{Apostol-97} is defined by the infinite product%
\begin{equation}
\eta(\tau)=e^{i(\pi/12)\tau}\prod\limits_{n=1}^{\infty}(1-e^{2\pi in\tau})\,.
\label{eta-infinite-product}%
\end{equation}
for $\mathrm{Im}\,\tau>0$.

At large $\mathrm{Im\,}\tau\rightarrow+\infty$%
\begin{equation}
\eta(\tau)\overset{\mathrm{Im\,}\tau\rightarrow+\infty}{\rightarrow}%
e^{i(\pi/12)\tau}\,. \label{eta-tau-large}%
\end{equation}
Dedekind $\eta$ function has the property%
\begin{equation}
\sqrt{z}\eta(iz)=\eta(iz^{-1})\, \label{eta-inv}%
\end{equation}
where the branch of $\sqrt{z}$ is chosen so that $\sqrt{z}$ is real for $z>0$.

Note that in the power series%
\begin{equation}
\left(  1-q\right)  ^{-\alpha}=\sum\limits_{k=0}^{\infty}\frac{\Gamma
(\alpha+k)}{k!\Gamma(\alpha)}q^{k}%
\end{equation}
all coefficients are positive at $\alpha>0$. Expanding the infinite product%
\begin{equation}
\prod\limits_{n=1}^{\infty}\left(  1-q^{n}\right)  ^{-\alpha}=\prod
\limits_{n=1}^{\infty}\left[  \sum\limits_{k=0}^{\infty}\frac{\Gamma
(\alpha+k)}{k!\Gamma(\alpha)}q^{nk}\right]  =\sum\limits_{n=0}^{\infty}%
A_{n}(\alpha)q^{n}\,, \label{1-min-q-expansion}%
\end{equation}
one finds that here all coefficients are also positive%
\begin{equation}
A_{n}(\alpha)>0\quad(\alpha>0)\,. \label{A-n-positive}%
\end{equation}
Applying this to Eq.~(\ref{eta-infinite-product}), we see that%
\begin{equation}
\left[  \eta(i\tau)\right]  ^{-\alpha}=\left[  e^{-(\pi/12)\tau}%
\prod\limits_{n=1}^{\infty}(1-e^{-2\pi n\tau})\right]  ^{-\alpha}%
=e^{(\pi/12)\alpha\tau}\sum\limits_{n=0}^{\infty}A_{n}(\alpha)e^{-2\pi n\tau
}\,. \label{eta-alpha-pos}%
\end{equation}

In the case $\alpha=1$ expansion (\ref{1-min-q-expansion}) simplifies to%
\begin{align}
\prod\limits_{n=1}^{\infty}\left(  1-q^{n}\right)  ^{-1}  &  =\prod
\limits_{n=1}^{\infty}\sum\limits_{k=0}^{\infty}q^{kn}=\sum\limits_{n_{1}%
=0}^{\infty}\sum\limits_{n_{2}=0}^{\infty}\sum\limits_{n_{3}=0}^{\infty}%
\ldots\delta_{n,n_{1}+2n_{2}+3n_{3}+\ldots}q^{n}\nonumber\\
&  =\sum\limits_{n=0}^{\infty}P(n)q^{n} \label{1-min-q-N-expansion}%
\end{align}
where $P(n)$%
\begin{equation}
P(n)=\sum\limits_{n_{1}=0}^{\infty}\sum\limits_{n_{2}=0}^{\infty}%
\sum\limits_{n_{3}=0}^{\infty}\ldots\delta_{n,n_{1}+2n_{2}+3n_{3}+\ldots}
\label{P-n-def}%
\end{equation}
is the number of partitions of the integer $n$, i.e. the number of
representations of $n$ in the form%
\begin{equation}
n=n_{1}+2n_{2}+3n_{3}+\ldots
\end{equation}
Taking an integer $m$ power of Eq.~(\ref{1-min-q-N-expansion})%
\begin{equation}
\left[  \prod\limits_{n=1}^{\infty}\left(  1-q^{n}\right)  ^{-1}\right]
^{m}=\left[  \sum\limits_{n=0}^{\infty}P(n)q^{n}\right]  ^{m}=\sum
\limits_{n=0}^{\infty}A_{n}(m)q^{n}\,,
\end{equation}
one can see that%
\begin{equation}
A_{n}(m)=\sum\limits_{n_{1}=0}^{\infty}\sum\limits_{n_{2}=0}^{\infty}%
\ldots\sum\limits_{n_{m}=0}^{\infty}\delta_{n,n_{1}+n_{2}+\ldots+n_{m}}%
P(n_{1})P(n_{2})\ldots P(n_{m})\,. \label{A-n-m-via-P-n}%
\end{equation}
This shows that all $A_{n}(m)$ are integer positive numbers if $m,n$ are integer.

\end{document}